\documentclass[fleqn,usenatbib]{mnras}

\usepackage{esvect}
\usepackage{amssymb}
\usepackage{esvect}
\usepackage{gensymb}

\usepackage[T1]{fontenc}
\usepackage{ae,aecompl}
\usepackage{graphicx}	
\usepackage{float}      
\usepackage{savesym}
\usepackage{amsmath}
\savesymbol{iint}
\usepackage{txfonts}
\restoresymbol{TXF}{iint}
\usepackage{amssymb}	

\ifdefined\theHchapter\else\newcommand\theHchapter{\Alph{chapter}}\fi
\ifdefined\theHsection\else\newcommand\theHsection{\Alph{section}}\fi

\title[Asymmetric star-formation due to Ram Pressure]{The better half - Asymmetric star-formation due to ram pressure in the EAGLE simulations}

\author[P. Troncoso-Iribarren et al.]{P. Troncoso-Iribarren$^{1}$, 
N. Padilla$^{2}$,
C. Santander $^{2}$,
C.D.P. Lagos $^{3,4,5}$,
\newauthor{D. Garc\'ia-Lambas$^{6,7}$,
S. Rodr\'iguez$^{6}$,
S. Contreras$^{2,8}$}
\\
$^{1}$ Escuela de Ingenier\'ia, Universidad Central de Chile, Avda. Francisco de Aguirre 0405, La Serena, Chile.\\
$^{2}$ Centro de Astro-Ingenier\'ia \& Instituto de Astrof\'isica, Pontificia Universidad Cat\'olica de Chile, Avda. Vicu\~na Mackenna 4860, \\ 782-0436 Macul, Santiago, Chile. \\
$^{3}$ International Centre for Radio Astronomy Research (ICRAR), M468, University of Western Australia, 35 Stirling Hwy, Crawley, \\WA 6009, Australia.\\
$^{4}$ ARC Centre of Excellence for All Sky Astrophysics in 3 Dimensions (ASTRO 3D).\\
$^{5}$ Cosmic Dawn Center (DAWN).\\
$^{6}$ Instituto de Astronom\'ia Te\'orica y Experimental, UNC-CONICET, C\'ordoba, X5000BGR, Argentina.\\
$^{7}$ Observatorio Astron\'omico de C\'ordoba, Universidad Nacional de C\'ordoba, X5000BGR, Argentina.\\
$^{8}$ Donostia International Physics Center (DIPC), Manuel Lardizabal Pasealekua 4, E-20018 Donostia, Basque Country, Spain
}

\date{Accepted 17 January 2020. Received 05 December; in original form 25 September}
\pubyear{2020}
\begin{document}
\label{firstpage}
\pagerange{\pageref{firstpage}--\pageref{lastpage}}
\maketitle
\begin{abstract}
We use the EAGLE simulations to study the effects of the intra-cluster medium (ICM) on the spatially resolved star-formation activity in galaxies.
We study three cases of galaxy asymmetry dividing each galaxy in two halves using the plane (i) perpendicular to the \texttt{velocity} direction, differentiating the galaxy part approaching to the cluster center, hereafter dubbed as the ``leading half'', and the opposite one ``trailing half'',
(ii) perpendicular to the \texttt{radial} position of the satellite to the centre of the cluster,
(iii) that maximizes the star-formation rate ($\rm SFR$) difference between the two halves.
For (i), we find an enhancement of the $\rm SFR$,  star formation efficiency ($\rm SFE$), and interstellar medium pressure in the leading half with respect to the trailing one and normal star-forming galaxies in the EAGLE simulation, and a clear overabundance of gas particles in their trailing.
These results suggest that ram pressure (RP) is boosting the star formation by gas compression in the leading half, and transporting the gas to the trailing half.  This effect is more pronounced in satellites of intermediate stellar masses $\rm 10^{9.5-10.5} M_{\sun}$, with gas masses above $\rm 10^{9} M_{\sun}$, and located within one virial radius or in the most massive clusters.
In (iii) we find an alignment between the velocity and the vector perpendicular to the plane that maximizes the $\rm SFR$ difference between the two halves. It suggests that finding this plane in real galaxies can provide insights into the velocity direction. 

\end{abstract}




\section{Introduction}

Galaxy clusters are important laboratories to study galaxy evolution because they allow us to explore the maximal effect of interactions and nurture on the evolution of galaxies. 
As galaxies plunge into the intra-cluster medium (ICM) of groups and clusters of galaxies, they experience different effects from their surrounding environment including ram pressure (RP) and tidal stripping.
These are believed to produce changes in the galaxy properties ranging from gas loss to enhanced star formation (SF) activity due to the increase of pressure acting on the disc of the galaxy \citep{kapferer09,steinhauser12,safarzadehloeb19}.
Since the seminal work of \citet{gunngott72}, RP is a well known physical mechanism thought to be a major driver behind the observed absence of spiral galaxies in the central regions of dense cluster environments. 

\citet{gunngott72} showed that RP is proportional to the density of the ICM times the relative velocity of the satellite galaxy and the brightest cluster galaxy.
RP may accelerate the star-formation rate ($\rm SFR$) of the galaxies residing in dense environments, thus prompting their transition from an active to a passive state \citep{gunngott72,moore96,jaffe15}.

In observations, the effect of RP has been studied in galaxies that are being gas-stripped in galaxy clusters, identified via their distorted morphologies \citep{smith10,mcpartland16}.
Historically, this identification has been performed via visual inspection of images of cluster members \citep{poggianti16,mcpartland16}, and in mock images of hydrodynamical simulations \citep{yun19}. It allows to determine their abundance and dependence with redshift.
Ideally, integral field spectroscopy (IFS) is used to map the spatially resolved emission of the stellar and gas components, its kinematics and re-construct their evolution.
The GASP (GAs Stripping Phenomena in galaxies with MUSE) survey performed an unprecedented IFS detailed study of local galaxies, in the redshift range 0.04-0.07 \citep{poggianti17,bellhouse19}.
This ESO large program uses 120 hours to push the state-of-art, observing one hundred local galaxies in the field and clusters to a level of detail much greater than can be done in larger surveys with IFS such as SAMI (Sydney-Australian-Astronomical-Observatory Multi-object IFS, \citet{croom12}) and MaNGA (Mapping Nearby Galaxies at APO, \citet{bundy15}).
Yet, the number of objects analyzed in the GASP and SAMI or MANGA survey differs by two orders of magnitude.

The advent of large scale surveys motivates to find other methods to study galaxy evolution and specifically RP in a statistical manner.
Particularly in view of photometric surveys such as SDSS (Sloan Digital Sky Survey, \citet{sdss09}), 
VST-ATLAS \citep{shanks15}, J-PAS \citep{jpas}, 
LSST \citep{lsst} in the near future or spectroscopic ones like GAMA \citep{baldry10},
DESI \citep{desi16}, BOSS \citep{boss13}, DEVILS \citep{davies18}, and WAVES \citep{waves}.
They permit to measure galaxies under the effect of RP in thousands of clusters at different redshifts.
Machine learning techniques have been used to classify galaxies according to their morphology \citep{banerji10} and could be used to identify galaxies undergoing strong transformation processes or RP effects, for example.

In this work, we propose a statistical method using the EAGLE \citep{eaglepaper} suite of hydrodynamical simulations immersed in a cosmological context to explore statistical ways to measure the RP effect on cluster galaxies and its consequences on other physical properties of galaxies.
The rationale of this approach is twofold.
We use hydrodynamical simulations because they allow to explore the spatially resolved properties and study galaxy evolution self-consistently. It permits to select large and statistically complete samples. 
We use the EAGLE galaxies because they follow the general scaling relations between stellar mass, $\rm SFR$, metallicity, and gas content, as it is described in detail in \cite{clau_fp}. 
They reproduce the observed galaxy colours \citep{trayford15}, the fraction of passive galaxies as a function of stellar mass \citep{furlong15}, and the slope of the spatially resolved $\rm SFR$- stellar-mass and mass-metallicity relations, down to kilo parsec scales \citep{trayford19}.

We analyze all the galaxies forming stars, above certain $\rm SFR$ or mass limit, identified as members of clusters and groups, and use the simplest approach of dividing each galaxy satellite into halves.
The one that faces the medium as it moves through the ICM, which we will refer to as the leading half, and the one facing the opposite way, the trailing half of the galaxy.
If the timescale for the effect of RP stripping on the SF activity is shorter than the dynamical timescale of the disc, the enhancement of the SF should be more prominent in the leading half, which will be subject to the effect of RP.
In this case we can expect to find different average SF properties or colors in the leading half with respect to the trailing one.

We apply this method to samples of simulated galaxies, in order to study the dependence of this effect with cluster and galaxy properties.
We aim to measure the property differences between the leading and trailing halves of the galaxies and study whether these differences are detectable with current and future instruments and facilities.

This work is organized as follows: in section \S \ref{sec2} we describe the simulation and methods.
Three cases for halving the galaxies are presented.
The first one is dubbed ``the \texttt{velocity cut}" because it uses the three dimensional velocity vector of satellites to define the leading and trailing half. 
In typical observations of large galaxy samples, it will be difficult to obtain this vector. 
It motivates us to present two cases of possible observational application based on the three dimensional position vector, of the center of potential, of satellites (section \S \ref{sec_methods}). 
With the current accuracy achieved in photometric redshifts of large scale surveys \citep{ascaso16}, we expect it possible.
The analysis of mock galaxies and projected images will be presented in a separate article.
In the first observationally driven case, we choose the most simplistic approach of studying the differences between the half that faces the center of the group/cluster, and compare it to the opposite one \citep{troncoso16b}.
In section \S \ref{sec_results}, we show the results for the first case.
In section \S \ref{sec_perrito}, we study our second observationally driven case, namely the \texttt{maximum anisotropy cut}, in which we divide the galaxy according to the plane that maximizes the $\rm SFR$ difference.
In the same section we analyze the differences between the two observationally driven cases.
In sections \S \ref{sec_dep_cluster}, \S \ref{sec_dep_gp}, \S \ref{sec_ms} we analyze the dependence of the first case with the cluster and galaxy properties, and compare it to main-sequence galaxies in the EAGLE simulation, respectively. 
Finally in section \S \ref{sec_discusion}, we use the three cases to discuss our findings and study the strength of the anisotropy on cluster and galaxy properties. In section \S \ref{sec_conclusions} our conclusions are presented.

\begin{figure*}
	\includegraphics[width=2.\columnwidth, trim=0cm 0cm 0cm 0cm]{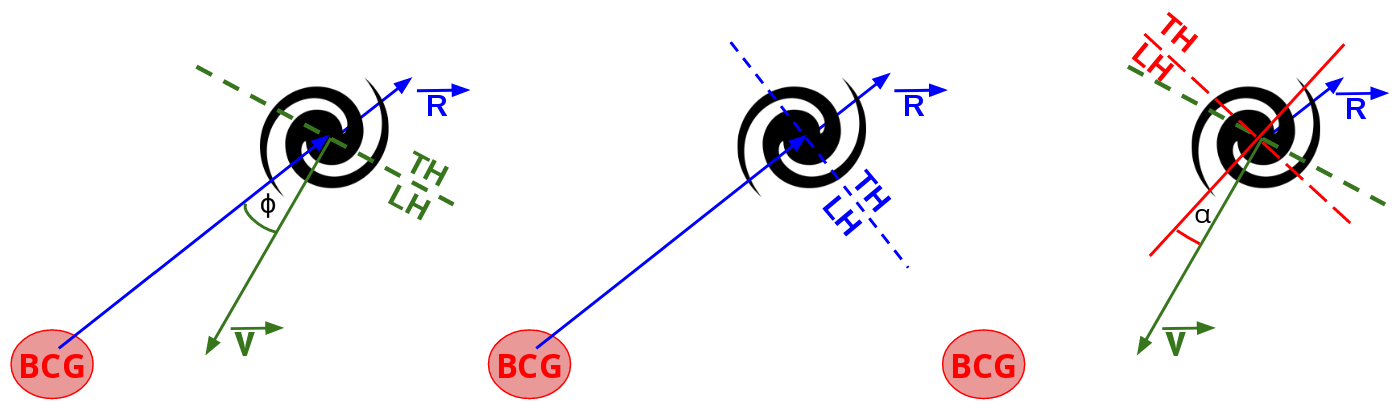}
    \caption{Cartoon of a galaxy cut in two parts according the three cases discussed in this work. 
    The green line indicates the three dimensional velocity direction of the galaxy. The dashed green line indicates the plane perpendicular to it, which defines the first case, using the \texttt{velocity cut}, schematized in the {\it left panel}.
    The blue line shows the three dimensional position of the center of mass of the galaxy with respect to the central galaxy. The dashed blue line is perpendicular to it and defines the plane to divide the galaxy in the \texttt{radial cut},  shown in the  {\it middle panel}. 
    The red dashed line represents the plane that maximizes the $\rm SFR$ difference between the two halves. This plane is searched for in every analyzed galaxy as it is described in section \ref{sec_perrito}, once it is found, it defines the halves of the \texttt{maximum anisotropy cut}.
    The red solid line is perpendicular to this plane. It is used to define the alignment angle, $\alpha$, with respect to the velocity direction of the galaxy.}
    \label{fig1_scheme}
\end{figure*}

\section{Simulation and Methods} \label{sec2}

\subsection{EAGLE simulation}
\label{sec_eagle} 

The EAGLE project \citep{eaglepaper} is a suite of hydrodynamic simulations immersed in a $\rm \Lambda CDM$ cosmology,
adopting the Planck Collaboration XVI (2014) cosmological parameters.
It uses up to seven billion baryon and dark matter particles per individual simulation to follow the physics of galaxy formation. 
These simulations were performed within the framework of the Virgo consortium and broadly reproduce the properties of the local Universe and galaxies as our Milky Way \citep{furlong15,trayford15,clau_fp}. 
For further details please see \cite{eaglepaper} and \cite{eaglepaper_det}.
The particle data and halo catalogues are publicly available in the EAGLE website\footnote{http://icc.dur.ac.uk/Eagle/database.php}.

In this work we use the RefL0100N1504 simulation, a periodic box of $\rm 100 cMpc$ of length and $\rm 1504^3$ gas and dark matter particles in its initial conditions.  
This simulation is the one with the largest volume in the EAGLE suite, and therefore provides the best statistical sample of galaxies in groups and clusters.
We select groups/clusters in the range of $\rm log_{10}( M/M_{\sun})=13.6-14.6$, concentrating on large satellites to avoid resolution problems, with typically thousands (hundreds) of stellar (gas) particles per galaxy.
The simulation suite was run with
a modified version of the GADGET-3 Smoothed Particle Hydrodynamics (SPH) code \citep{springel05}. These modifications are collectively dubbed {\texttt ANARCHY} \citep{schaller15}.
It was run with a suite of subgrid models including radiative cooling and photoheating \citep{wiersma09a}, stellar evolution and chemical enrichment \citep{wiersma09b} taking into account energy feedback from the stars \citep{dallavecchiaschaye12}, black hole growth and active galactic nuclei \citep{rosasguevara15}. Following the SF law of \cite{schayedallavecchia08}, the $\rm SFR$ of each gas particle is calculated with the Kennicutt-Schmidt (KS) law \citep{kslaw} based on a pressure scheme 
\begin{equation}
\rm SFR_i = m_{i} A \left[1~\rm \frac{M_{\sun}}{pc^{2}}\right]^{-n} \left ({\frac{\gamma}{G}} \rm f_{\rm g} \bar P_i\right )^{(n-1)/2}, \label{eq_sfr}
\end{equation}
where $\rm m_{i}$ is the mass of the gas particle, $\gamma =5/3$ is the ratio of specific heats, $G$ is the gravitational constant, $f_{\rm g}$ is the mass fraction in gas (assumed to be unity), and $P_{\rm i}$ is the entropy-weighted average pressure. 
$A$ and $n$ are fixed to the observational results of the KS law, $A=1.515\times 10^{-4} \rm M_{\sun}/yr/kpc^2$ and $n=1.4$.

In this {\texttt ANARCHY} version of pressure-entropy SPH, each gas particle i carries its (pseudo) entropy $S_i$, which is used to solve the hydrodynamics of each particle (see equation 4 in \citet{eaglePDR}), and calculate their entropy-weighted average pressure and entropy-weighted average density following,
\begin{equation}
    \bar P_i= S_i \left( \frac{1}{S_i^{1/\gamma}} \sum_{j=1}^N m_j S_j^{1/\gamma} W_{ij} (h_i) \right)^\gamma  \equiv S_i \bar \rho^\gamma,
\end{equation}

where $W_{ij} (h_i)$ is the value of the kernel at that location. EAGLE uses the $C_2$ kernel of \citet{eaglekernel}. The total $\rm SFR$ of the galaxy is the sum of the $\rm SFRs$ of individual gas particles.

We normalize the pressure, $P/k$, with the Boltzmann constant and use the conversion factor $ 1.38\times 10^{-16} [dyn/cm^2]$ or $10^{-17} [N/m^2]$ to compare with observational measurements. 
We calculate the pressure of each particle, without weighting by the entropy, using the equation of state $ P_i= S_i \rho_i^\gamma$, where $\rho$ is the standard SPH density.

To properly model SF, EAGLE introduces density and metallicity thresholds for certain cases, below which SF is not feasible or simply to avoid modeling cases that are non-resolved by the limits of the simulation.
This prevents unrealistic cases such as, hot/metal rich gas forming stars or the formation of spurious stars at high-redshift, when the mean density of the Universe is similar to the critical density of SF. 
Gas can be converted into stars only if it manages to cool down and reach high densities. Because the efficiency of gas is a strong function of the gas density and metallicity, \citet{schaye04} introduced a threshold density above which stars form that is metallicity-dependent,

\begin{equation}
\rm n_{\rm H}(Z)= 0.1\, \left ({\frac{\rm Z}{0.002}}\right )^{-0.64} ,\label{eq_Zsfr}
\end{equation}
where $\rm Z$ is the metallicity of the gas.\\

\begin{figure*}
\includegraphics[width=0.6\columnwidth, trim=0cm 0cm 0cm 0cm]{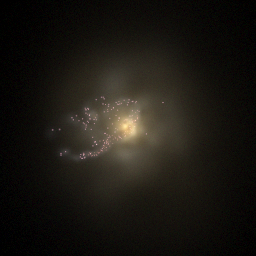}
\includegraphics[width=0.6\columnwidth, trim=0cm 0cm 0cm 0cm]{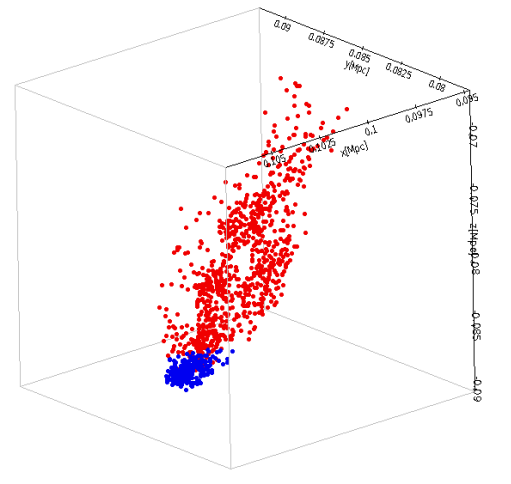}
\includegraphics[width=0.8\columnwidth, trim=0cm 0cm 0cm 0cm]{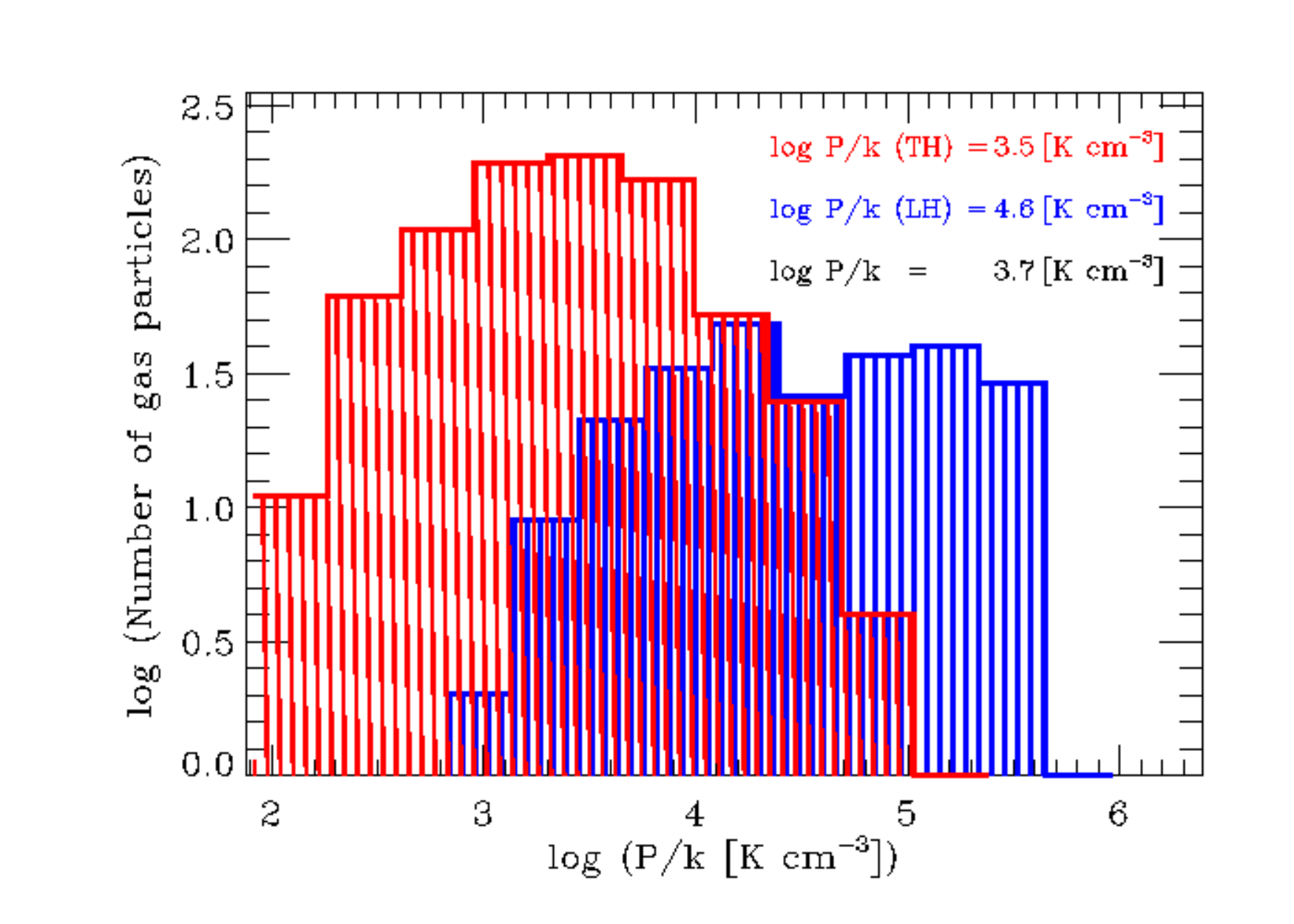}
\caption{Analysis of the individual EAGLE galaxy with ID 6082966 residing in  group 19 (subgroupnumber 11),  falling into its central galaxy with an angle of 110 degrees.
{\it Left:} face-on image composed of the emission in the SDSS bands u, g, and r, derived using the radiative transfer simulations of SKIRT \citep{trayford17,baes11}.
The field of view is 0.06 proper Mpc (pMpc).
Image credits to the EAGLE consortium, taken from the public database \citep{mcalpine16}. {\it Middle:} three dimensional view of its gas particles;  blue shows the position of the leading half while  red, the trailing half. Axes are in pMpc. The orientation of this cube is different to the image in the left panel.
{\it Right:} Logarithmic pressure histogram of the gas particles in the leading (blue) and trailing halves (red). }
\label{fig_pressure}
\end{figure*}

\subsection{The galaxy sample}
\label{sec_sample}

\subsubsection{The satellite sample} \label{ssec_satellite}
We consider all galaxies of the largest simulation box of the suite RefL0100N1504 that reside 
in clusters/groups of mass greater than $\rm 10^{13.6} M_{\odot}$ in snapshot 28, corresponding to $z=0$. 
These groups are found using the multistage procedure, based on a friends of friends ($\rm FOF$) algorithm, described in \cite{eaglepaper}.

For the mass scale of the clusters and groups, we use an overdensity of $\rm M_{200}^{crit}$ as in \cite{eaglepaper_det}.  $\rm M_{200}^{crit}$ is the mass enclosed in a radius with a density of two hundred times the critical density.
The mass limit $\rm M_{200}^{crit} > 10^{13.6} M_{\odot}$ selects the first 25 most massive haloes of the simulation, hosting 3062 galaxies altogether of any stellar mass.

We exclude of this analysis the halo with GroupNumber=17 because the position difference between the center of mass and the maximum of the potential, normalized to its virial radius, is one order of magnitude larger than all the other haloes. This kind of disparity is usually associated with merging galaxy clusters.
This suggests that it is not in equilibrium.  
We discard 117 galaxies by excluding this cluster.

Galaxies are identified using the algorithm $\rm SUBFIND$ \citep{springel01,dolag09}. 
It is used to identify local over-dense self-bound substructures or sub-haloes, within the full particle distribution of $\rm FOF$ haloes.
Here, we consider all the particles found by the algorithm and not constrain it to certain spherical aperture. It is in contrast with most of the EAGLE publications, which commonly used 30 pkpc \citep{trayford19}.

Since we are interested in the analysis of star-forming galaxies, we select the ones with a mass of star-forming gas, $\rm Mass (SF_{gas} )> 0$, reducing our sample to 513 galaxies.
Additionally, we only analyze  galaxies that are well suited for statistics, i.e. with more than one hundred particles in their stellar and gas components $\rm (N_{gas}>100$ and $\rm N_{stars}>100)$. 
This reduces the sample to 227 galaxies. By selecting the satellites that reside within one virial radius, 16$\%$ of the total satellite galaxies that form stars in the simulation at z=0 remain in our sample (80 galaxies). 
This condition might be too conservative; relaxing the threshold to 50 particles, enlarges the sample in 20$\%$ but does not change the final results and conclusions and hence we preserve the present choices for the definition of our sample.
Within one virial radius, there is only one galaxy with stellar mass  $< \rm 10^{8.85} M_{\sun}$;  we decided to exclude it and make this the stellar mass cut for our final sample selection, i.e 79 (211) satellites at $r<$ 1(3)$\times \, r_{vir}$, respectively.
Our final sample spans a wide range of stellar masses, $\rm  8.85< log_{10}\,M_* [M_{\sun}] <11.5$, star formation rates, $ \rm -1.3 < log_{10}\,SFR [M_\odot/yr]< 1.1$, gas masses $\rm 8.5 < log_{10}\,M_{gas} [M_{\sun}]<10.5$, and masses of star-forming gas $\rm 8.5 < log_{10}\,M_{SFgas} [M_{\sun}] <10.5 $, metallicities, and gas fractions (see fig. \ref{fig_sfrenh_gp}).

\subsubsection{The control sample} \label{sec_controlsample}

 In order to compare the properties of our satellites with the galaxies that are non affected by environmental process, we select a control sample of central and isolated galaxies.
 We ran a search for galaxies in the snapshot 28 that fulfill the conditions
 i) it is a central galaxy (SubGroupNumber=0) and ii) it is consider as a field galaxy, $\rm 3 \times 10^{11}< M_{halo} [M_\odot]> 2 \times 10^{12}$, 
 iii) it does not have neighbor galaxies of stellar masses greater than $\rm  10^{9}[M_\odot]$. We found 3182 galaxies. 

 From those, we selected the ones with properties similar to our satellites sample, it is $\rm log M_*>8.85$, 
 $\rm N_{gas} >100$, $\rm -1.3 < log_{10}\, SFR [M_\odot/yr]< 1.1$,
 $\rm log\, M_{gas}>8.5$, and $\rm\,log M_{SFgas}>8.5$, 
 ending up into 2191 galaxies.
 We remark that all galaxies in this sample have $\rm sSFR > 0.01 [Gyr^{-1}]$. Hence they follow the main-sequence criterion defined in \citet{furlong15}.
 
 We define a second control sample, hereafter dubbed "main-sequence sample", following
 \citet{furlong15}, we select the galaxies with $\rm sSFR [Gyr^{-1}] > 0.01 $,  $log\,M_*[M_\odot] > 9 $, and imposing the condition of being a central galaxy $SubGroupNumber = 0$, . 
 This sample contains 6340 galaxies.
 It will be used in sections \S \ref{sec_ms}, and \S \ref{sec_discusion} to connect our results with observational works.

\subsection{Methods} \label{sec_methods}

The EAGLE consortium provides a wide variety of physical properties for each gas, dark matter, and stellar particle \citep{eaglepaper_det}. For the positions, velocities, $\rm SFR$s, entropies, densities, metallicities, and element abundances of gas particles, and the positions, velocities, metallicities, element abundances, and ages of stellar particles.

\subsubsection{Halving the galaxies}
\label{sec_mns}

In the following, three different cases are studied depending on the plane that divides the galaxy. For all cases the three-dimensional structure of the galaxy is analyzed. In a forthcoming study we will study projection effects and EAGLE mock images \citep{trayford17} available in the public database \citep{mcalpine16}.

Firstly, we define the \texttt{velocity cut}, which refers to the measurements that can be performed when the three-dimensional velocity of the satellite's center of potential is known.
In this case, the galaxy is halved according to the plane that contains its centre of potential, and that is normal to its velocity vector relative to the cluster center of mass.

It can be expected that this plane will maximize the $\rm SFR$ difference between the two halves as it is the direction in which RP should be maximal.
Yet, the three-dimensional velocity vector is not an observable quantity.

The second case is dubbed \texttt{radial cut} because it refers to the most simplified version that could be performed in current observations of galaxies in clusters.
The galaxy is divided with respect to the plane normal to its three-dimensional position vector with respect to the cluster center of mass. 
We remark that in this case the three-dimensional position vector of the cluster member is used. 
In current observational studies only the two dimensional information is used and this cut can be applied to upcoming data with precise photometric redshifts of each cluster member. 
These two cases are explained and discussed in \cite{troncoso16b}.

We define a third case using the plane that maximizes the $\rm SFR$ difference; a 2-dimensional version of this cut would also be applicable to observational data.
We search for the plane that maximizes the $\rm SFR$ difference between two galaxy sides and explore whether this maximum difference correlates with the velocity vector. This last case is dubbed \texttt{maximum anisotropy cut} and the results are presented in section \ref{sec_perrito}.
 Fig.~\ref{fig1_scheme} shows the cartoon representation of the three cases.

We define the \texttt{falling angle} as the angle between the three-dimensional position and the velocity vector of the galaxy, both relative to the central galaxy of each cluster
\begin{equation}
  \rm  cos (180 - \phi) = \frac{(\vec{r}_{\rm CM}-\vec{r}_{\rm CG}) \cdot (\vec{v}_{\rm CM}-\vec{v}_{\rm CG})}{ ||(\vec{r}_{\rm CM}-\vec{r}_{\rm CG})|| \cdot ||(\vec{v}_{\rm CM}-\vec{v}_{\rm CG})||} .
	\label{eq_angle}
\end{equation}
This angle contains information about the ellipticity of the orbit.

\subsubsection{Measuring characteristic properties of each galaxy half}
\label{sec_newobs}

Visually inspecting the EAGLE galaxy shown in Fig. \ref{fig_pressure}, we realize that summing a property over all particles in each half may bias the difference between  halves because one of them may be more massive or contain more particles than the other one (the adopted center is that of the gravitational potential).
Hence, the intrinsic asymmetries between the two halves could bias the $\rm SFR$, density, pressure, etc.

In section \ref{sec_percentages}, we discuss the implications of directly integrating over each galaxy half, considering different numbers of particles. 
Hereafter, we define observables that are independent of the intrinsic mass asymmetries between the two halves. 

The $\rm SFR$ enhancement or percentage excess of the leading half with respect to the trailing half is defined as 
\begin{equation}
\rm \delta SFR^{mw}_{E} \equiv \frac{SFR^{mw} (LH) - SFR^{mw} (TH)}{SFR^{mw}_{\rm total}},
\label{mw_sfre}
\end{equation}
where $\rm SFR^{mw}(LH)$, $\rm SFR^{mw}(TH)$ are the mass-weighted average $\rm SFR$ of the gas particles in the leading and trailing halves, respectively.
The normalization term $\rm SFR^{mw}_{total}$ corresponds to the gas mass-weighted average $\rm SFR$ of all gas particles in the galaxy.

Thus, this quantity is unaffected by the mass difference between the two halves.
\\
We also calculate the median enhancement, which is defined as the difference between the median values of each half, normalized to the median value of the complete galaxy. This quantity is insensitive to the difference in the number of particles between the two halves.
For example, the median pressure enhancement of the gas particles is defined as follows,
\begin{equation}
\rm \delta P^m_{E} \equiv \frac{P^m (LH) - P^m (TH)}{P^m_{total}},
\end{equation}
where $\rm P^m$(LH), $\rm P^m$(TH) are the medians value of the pressure of the gas particles in the leading and trailing halves, respectively. $\rm P^m_{total}$ is the median value overall gas particles in the satellite.
Analogously, volume weighted average quantities can be measured in order to take into account the volume represented by each particle. 
For a particular particle property, these three quantities can be constructed; i.e. for the metallicity, we calculate the mass-weighted average enhancement, $\rm \delta Z^{mw}_{E}$, the median enhancement $\rm \delta Z^m_{E}$, and the volume-weighted average enhancement $\rm \delta Z^V_{E}$.
These quantities might differ if extreme asymmetries between the two halves are present.
If the number, mass, and volume of the particles in each half are similar, then the three quantities described above converge to similar values.
In the particular case of the SF (see equation \ref{eq_sfr}) it is, intrinsically, an entropy-weighted quantity because it is  based on the entropy-weighted pressure of each particle.
The quantity $\rm \delta SFR^{mw}_E$ is then weighted by the entropy and gas mass of the particles.

We define the enhancement of the star-formation efficiency ($\rm SFE$) as,
\begin{equation}
\rm \delta SFE  \equiv \frac{SFE^{mw}(LH) - SFE^{mw}(TH)}{SFE^{mw}_{\rm total}},
\end{equation}
where, $\rm SFE^{mw}= \frac{ \sum m_{gas,i} \dot\, (SFR_i/m_{gas,i})}{\sum m_{gas,i}} $, is the gas mass-weighted average of the $\rm SFE$. 
The numerator of the $\rm SFE$ is the sum over the individual $\rm SFR_i$, while the denominator is the sum of the gas mass of the particles. Both depend on the number of particles, but the ratio between them is independent of the resolution.

Another way to formulate the  enhancement of a property is the logarithmic ratio or excess between the two halves, for example 
\begin{equation}
\rm SFR^m_{LRE} \equiv log_{10} \frac{SFR^m(LH)}{SFR^m(TH)},
\label{eq_lre}
\end{equation}
where $\rm SFR^m(LH)$, $\rm SFR^m(TH)$ are the median $\rm SFR$ of the leading and trailing halves, respectively.
The analogous quantity for the gas mass-weighted average $\rm SFE$ ,
\begin{equation}
\rm SFE^{mw}_{RE} \equiv log_{10} \frac{SFE^{mw}(LH)}{SFE^{mw}(TH)}.
\label{eq_re}
\end{equation}

Finally, the integrated quantities, representing the global $\rm SFR$, pressure, age of each galaxy are calculated simply adding up all the individual values. 
With the exception of $\rm SFE$, the aforementioned quantities can differ from studies of the EAGLE team because they typically use an aperture of $\rm 30 \rm pkpc$ \citep{mcalpine16}, which corresponds to roughly an $\rm R_{80}$ Petrosian aperture that is a good approach for direct comparison with observations \citep{eaglepaper}.
In the case of the $\rm SFR$, more than $85\%$ of the galaxies give similar results (below 10$\%$ of difference) for $\rm SFR$ estimators suggesting that few particles are further than $\rm 30\, pkpc$ of the galaxy center of potential. 
Although those particles are not representative of the median, they record important information of the galaxy transformation processes occurring in situ and galaxy deformations.
For this reason, we suggest for future observational works to use the full photometry, building a surface brightness light profile for every galaxy and to measure the photometry in a radius such that it reaches the sky brightness level.

Another representative global value is the average over all particles in the galaxy (independent of the number of particles describing each galaxy).
For example, the average pressure of the galaxy is two orders of magnitude smaller than the pressure obtained from adding the individual pressure of each particle over all the galaxy. The former might be associated to the restoring force or self-gravity of the galaxy, referred to as anchoring force or $ \rm \Pi_{gal}$ in the observational work of  GASP  \citep{yara18,bellhouse19}.

\subsubsection{Cluster dynamical state}
\label{sec_dynamical_state}
To determine the level of dynamical relaxation  of each cluster, 
we measure the position shift between the center of mass and the minimum of the potential, normalized to its virial radius, 
which we refer to as relaxation index.
We divide our group sample into two: those with relaxation index above and below the median index, which we referred to as relaxed and less relaxed clusters.
We will study whether this index plays a role in the effect of the ICM on SF.
The median cluster masses of the relaxed clusters is $\rm log_{10}\,M\,[M_{\sun}]= 13.86^{+0.05}_{-0.03}$, while the less relaxed are more massive, $\rm log_{10}\,M\,[M_{\sun}]=14.07^{+0.14}_{-0.07}$.

\begin{figure*}
\includegraphics[width=2.\columnwidth, trim=4cm 17.5cm 3cm 6.5cm]{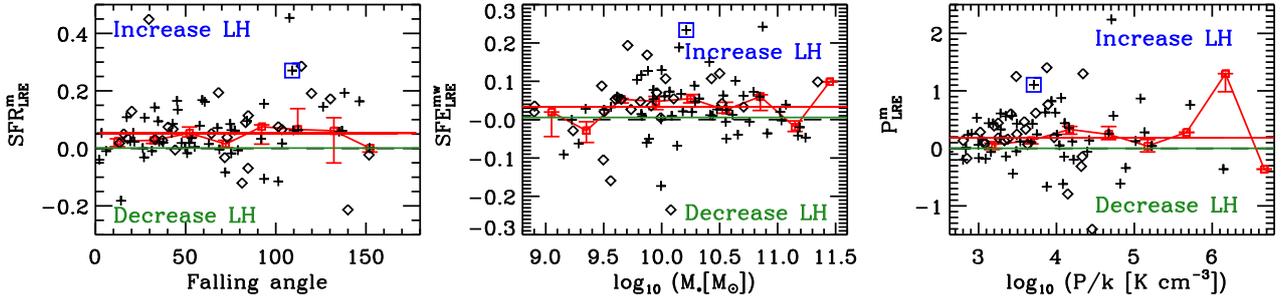}

\caption{Logarithmic ratio between leading and trailing halves of the median $\rm SFR$, gas mass-weighted average $\rm SFE$, and median pressure as a function of the falling angle, stellar mass and $\rm SFR$-weighted pressure. 
Galaxies within the 25 most massive groups, within one virial radius and  with more than one hundred gas particles are shown.
The halves were defined dividing the galaxies by the plane perpendicular to the velocity vector at their center of mass, with respect to the central galaxy.
The falling angle is defined in Eq. \ref{eq_angle}. 
In each panel, the solid green line shows the median value of the control sample.
The {\it left panel} shows the logarithmic ratio between the median $\rm SFR$ of the leading and trailing half as a function of the falling.
The {\it middle panel} shows the logarithmic ratio of the gas mass-weighted average $\rm SFE$ of the leading and trailing half as a function of the stellar mass.
The {\it right panel} shows the analogous quantity for the mean pressure as a function of the $\rm SFR$-weighted average pressure of the galaxy.
The black dashed line shows the case with equal pressure or $\rm SFR$ in both galaxy sides.
Red line shows the median of all galaxies.
Filled red squares are the medians at each bin, while the errors are the $1-\sigma$ scatter.
Diamonds show galaxies residing in relaxed clusters, while crosses in less relaxed clusters (see section \ref{sec_dynamical_state}).
The galaxy ID 6082966 that is shown in Figure \ref{fig_pressure} has been marked in a blue square in every panel.}
\label{fig_main}
\end{figure*}

\subsubsection{Jellyfish classification} \label{sec_jellyfish}
To discern if certain galaxy correspond to the jellyfish classification, we visually inspected in three dimensions their gas and stellar particles, and mock images available in the EAGLE database. 
A galaxy is classified as jellyfish if the star-forming gas particles are located in a preferential direction with respect to its center of potential or stellar content, i.e. the gas component presents extended tails showing a clear overabundance of gas particles in a preferential side. Following the criteria described in section \S 3 of \citet{yun19}, we search for asymmetric gas distributions or portions of asymmetric gas (Gas tails/wakes).

We also use the former defined quantities, Eq.\ref{eq_lre}, and Eq.\ref{eq_re} to rank our galaxy sample by the level of asymmetry.

\section{Results} \label{sec_results}

In this section we study the differences in the properties associated to the gas and stellar particles between the leading and trailing halves.
The $\rm SFR$, density, and pressure are properties inherent to the gas particles. 
Metallicities and oxygen abundances can be measured in both the stellar and gas particles, the former refers to the fraction of mass in elements heavier than Helium associated to each particle, while the latter corresponds to the ratio between the Oxygen and Hydrogen abundance.
Furthermore, the scale factor at which stellar particles were formed is stored, and we use this information to study the ages of the stellar populations.

We analyze galaxies that reside up to three virial radius of the cluster and have a minimum of one hundred stellar and gas particles describing the stellar and gas components, separately, as described in section \S \ref{sec_sample}. 
The latter condition ensures that each half is traced with reasonable statistics.

In the following subsections, the results for the gas and stellar particles are presented. Individual symbols in the plots shows the properties of individual satellites within one virial radius; the trends for satellites located within three virial radius are also analyzed.

We study the  differences of properties in the leading and trailing halves, firstly as a function of the {\it falling angle} defined in section \ref{sec_sample}, secondly as a function of  cluster properties (mass and radius), and finally as a function of the galaxy global properties (stellar mass, $\rm SFR$, mass of gas).
Hereafter, in all instances, except in sections \ref{sec_perrito} and \ref{sec_dep_cluster}, the results of the first case are reported. 
The reported values correspond to the median of the sample or at each bin.
The asymmetric error bar indicates the error with respect to the median, i.e. the upper and lower one sigma percentile normalized by the square root of the number of galaxies, $+\sigma/\sqrt{N}$, $-\sigma/\sqrt{N}$, respectively.
The only exception occurs in section \S \ref{sec_ms}, in which we report the one sigma percentile at each bin.
The properties of the control sample are reported in table \ref{tab_1}, and the median value is shown in each figure with a green solid line.

\subsection{Gas particles}

From this point on, we analyze only the gas particles which have $\rm SFR>0$; this way we select the star-forming gas that traces the galaxy disc.

For the galaxy sample described in section \ref{sec_sample}, the median of the number of gas particles describing each galaxy is $\rm 519^{+81}_{-39}$, while the hydrogen number density corresponds to $\rm n_{H} = 1.00^{+0.31}_{-0.08} [cm^{-3}]$, errors indicate the error on the median.
The gas particles in the simulation follow an ideal-gas equation of state (EOS), only gas which is deemed to be star-forming is placed on a artificial EOS that accounts for the sub-grid physics which takes place at high-density regions. It ensures that the ISM pressure increases with density.
Hence, the temperature is artificial and will not be used in the following analysis.

Figure \ref{fig_pressure} presents the spatially resolved properties of the individual EAGLE galaxy ID6082966.
The left panel of Figure \ref{fig_pressure} shows a face-on image, composite in the u, g, and r filters of the SDSS. The image was taken from the public data release \citep{mcalpine16} and created with the radiative transfer code of \citet{trayford17}.
The middle panel of Fig.~\ref{fig_pressure} shows the three-dimensional view of the gas particles divided in two parts according to the plane perpendicular to the velocity vector with respect to the central galaxy.
Blue diamonds indicate particles of the leading half, while the red dots correspond to the trailing one. Each side of the cube is in proper megaparsecs (pMpc).
The right panel of Fig.~\ref{fig_pressure} shows the logarithmic counts of the gas particles in logarithmic pressure bins for the leading (blue) and trailing (red) half. It shows an increase in pressure of the leading half (blue histogram) with respect to the trailing one (red histogram), by a factor of $\sim 6$.
The pressure of the leading half might be increasing due to the compression against the ICM. The middle and right panels evidence that the trailing half of this galaxy contains more gas particles than the leading half.
In section \ref{sec_percentages}, we study whether this is a particular case or a typical behaviour of cluster galaxies.

In Figure \ref{fig_main}, we show quantitative measurements of the difference in properties for all the galaxies in our sample defined in section \S \ref{sec_sample}.
The {\it left} panel shows the logarithmic ratio enhancement of the median $\rm SFR$ as a function of the falling angle.
The {\it middle} panel presents the logarithmic ratio enhancement of the gas mass-weighted average $\rm SFE$ as a function of the stellar mass.
The {\it right} panel shows the logarithmic ratio enhancement of the median pressure as a function of the $\rm SFR$-weighted pressure.
In each panel, the red line indicates the median of the logarithmic ratio enhancement of the median $\rm SFR$, gas mass-weighted average $\rm SFE$, and median pressure enhancement, 
$\rm \langle SFR_{LRE}^m \rangle = 0.05^{+0.01}_{-0.01}$, 
$\rm \langle SFE_{LRE}^m \rangle = 0.03^{+0.01}_{-0.01}$, 
$\rm \langle P_{LRE}^m \rangle = 0.18^{+0.05}_{-0.04}$, respectively.
The error bars indicate the error on the medians.
In each panel, the solid green line shows the median value of the control sample. This values are also reported in table \ref{tab_1}.

There is a positive enhancement of the satellites with respect to the control sample. It indicates that the median $\rm SFR$, gas mass-weighted average $\rm SFE$, and median pressure are higher in the leading half with respect to the trailing half. 

Similar median values for the satellite sample, described in section \ref{sec_sample}, are obtained if the differences of the mass-weighted average, or mean $\rm SFR$ are considered instead.
For example, the median of the mass-weighted average and median $\rm SFR$ are 
$ \rm \langle \delta SFR_E^{mw} \rangle  = 0.12^{+0.02}_{-0.02}$, 
$ \rm \langle \delta SFR_E^{m} \rangle = 0.12^{+0.03}_{-0.02}$, respectively. 
Analogously, the median enhancement of the $\rm SFR$-weighted average and mean pressure are 
$\rm \langle \delta P_E^{sw} \rangle = 0.40^{+0.08}_{-0.09}$,
$\rm \langle \delta P_E^{m} \rangle = 0.43^{+0.14}_{-0.09}$,
respectively.
The median pressure enhancement is strong with a median of $\rm 43 \%$ higher with respect to the median pressure of the galaxy, while the $\rm SFR$ increases only by $\rm 12\%$ with respect to the median $\rm SFR$ of the full galaxy.
These differences and logarithmic ratio enhancements are reported, respectively, in the columns 4-6, and 7-8 of Table \ref{tab_1}. We observe a linear correspondence between the individual measurements of the enhancement of the pressure and $\rm SFR$. It occurs because in the simulation both quantities are calculated as a function of the entropy-weighted average pressure (see Eq. \ref{eq_sfr}).

Galaxies falling into the cluster with a falling angle around $\rm \phi \gtrsim 90$ show the largest differences between leading and trailing halves (see left panel of Figure \ref{fig_main}).
The gas particles of the leading half are more compressed and are more efficient at forming stars than the ones of the trailing half.
In extreme cases, as the one of the galaxy ID 16300531 with $ \rm \langle \delta P_m^E \rangle =11.6$ and $\rm \langle \delta SFR_m^E \rangle =1.5$,
the difference between the median pressure of both galaxy sides can reach up to a factor of twenty, while the median $\rm SFR$ of the leading half is three times higher than the one of the trailing half.

By visually inspecting the EAGLE mock images of galaxies with $\rm SFR_{LRE}^m \gtrsim 0.2$, namely ID 6082966 and ID 10751313 ($\phi \gtrsim 90$), and ID 16300531 ($\phi \gtrsim 30$) we found that they correspond to the so-called jellyfish galaxies (see Figure \ref{fig_jellyfishEAGLE}).

These results, which are summarized in Figure \ref{fig_main} and Table \ref{tab_1}, suggest that RP compresses the gas, increasing the pressure, and boosting the $\rm SF$ of the leading half with respect to the trailing half.
\\
In Figure \ref{fig_main}, the diamonds indicate the galaxies residing in relaxed clusters, while the crosses show the galaxies of the less relaxed clusters. 
The median $\rm SFR$ enhancement for galaxies residing in relaxed clusters is  $\rm \langle \delta SFR_m^E \rangle =0.12^{+0.05}_{-0.04}$ and for the less relaxed, $\rm \langle \delta SFR_m^E \rangle =0.12^{+0.03}_{-0.02}$.
Their similarity confirms that the enhancement is independent of the relaxation classification described in section \ref{sec_dynamical_state}.\\

\begin{table*}
\begin{tabular}{c|c|c|c|c|c|c|c|c|c|}
\hline
&Sample  & $ \rm N_{gal}$   &$\rm N_{gas}$ &$\rm \delta SFR^m_E$  &$\rm \delta P^m_E$  &$\rm \Delta SFE$  &$\rm SFR^m_{LRE}$ &$\rm P^m_{LRE}$ &$\rm \Delta SFE_{LRE}$ \\
\hline
&     &   &   &  &$\rm r/r_{200} \le 1$ &\\
\hline
   &Satellites &      79   &$\rm 519^{+  81 }_{- 39}$   &$\rm 0.12^{+ 0.03}_{-0.02}$   &$\rm 0.43^{+ 0.14}_{-0.09}$   &$\rm 0.20^{+ 0.05}_{-0.03}$   &$\rm 0.05^{+ 0.01}_{-0.01}$   &$\rm 0.18^{+ 0.05}_{-0.04}$   &$\rm 0.03^{+ 0.01}_{-0.01}$\\
    \hline
    &R &      25   &$\rm 603^{+  37 }_{- 43}$   &$\rm 0.12^{+ 0.03}_{-0.02}$   &$\rm 0.44^{+ 0.14}_{-0.09}$   &$\rm 0.28^{+ 0.04}_{-0.04}$   &$\rm 0.05^{+ 0.01}_{-0.01}$   &$\rm 0.19^{+ 0.05}_{-0.04}$   &$\rm 0.05^{+ 0.01}_{-0.01}$\\
   &LR &      54   &$\rm 506^{+  86 }_{- 39}$   &$\rm 0.12^{+ 0.03}_{-0.02}$   &$\rm 0.40^{+ 0.14}_{-0.09}$   &$\rm 0.10^{+ 0.06}_{-0.03}$   &$\rm 0.05^{+ 0.01}_{-0.01}$   &$\rm 0.17^{+ 0.05}_{-0.04}$   &$\rm 0.02^{+ 0.01}_{-0.01}$\\
\hline
&     &   &  & &$\rm r/r_{200} \le 3$ & \\
\hline  
  &Satellites &     211   &$\rm 436^{+  46 }_{- 19}$   &$\rm 0.05^{+ 0.01}_{-0.01}$   &$\rm 0.23^{+ 0.06}_{-0.05}$   &$\rm 0.08^{+ 0.03}_{-0.02}$   &$\rm 0.02^{+ 0.01}_{-0.01}$   &$\rm 0.10^{+ 0.02}_{-0.02}$   &$\rm 0.02^{+ 0.01}_{-0.01}$\\
    \hline
    &R &      56   &$\rm 418^{+  27 }_{- 18}$   &$\rm 0.06^{+ 0.01}_{-0.01}$   &$\rm 0.32^{+ 0.05}_{-0.05}$   &$\rm 0.20^{+ 0.03}_{-0.03}$   &$\rm 0.03^{+ 0.01}_{-0.01}$   &$\rm 0.14^{+ 0.03}_{-0.02}$   &$\rm 0.03^{+ 0.01}_{-0.01}$\\
   &LR &     155   &$\rm 449^{+  53 }_{- 20}$   &$\rm 0.04^{+ 0.01}_{-0.01}$   &$\rm 0.15^{+ 0.06}_{-0.04}$   &$\rm 0.07^{+ 0.03}_{-0.02}$   &$\rm 0.02^{+ 0.01}_{-0.01}$   &$\rm 0.06^{+ 0.03}_{-0.02}$   &$\rm 0.01^{+ 0.01}_{-0.01}$\\
   \hline
   &Control   & 2191   &$\rm 209^{+   6 }_{-  3}$   &$\rm0.002^{+0.003}_{-0.003}$   &$\rm0.003^{+0.010}_{-0.011}$   &$\rm0.026^{+0.018}_{-0.018}$   &$\rm0.001^{+0.001}_{-0.001}$   &$\rm0.001^{+0.004}_{-0.005}$   &$\rm0.006^{+0.004}_{-0.004}$\\
  \hline
\end{tabular}
\caption{Gas properties of our satellite and control sample. Errors indicate the one sigma percentile of each galaxy sample. Notes. Column 1, sample name; Col. 2, number of galaxies in the sample; Col. 3, number of gas particles; Col., 4, 5, median $\rm SFR$ and pressure enhancement; Col., 6, Difference between the gas mass-weighted $\rm SFE$ of the leading minus the trailing half; Col., 7, 8, median $\rm SFR$ and pressure ratio enhancement; Col. 9, logaritmic ratio of the leading and trailing gas mass-weighted $\rm SFE$.}
\label{tab_1}
\end{table*}

\begin{figure}
\includegraphics[width=\columnwidth, trim=3.5cm 13.5cm 12cm 3cm]{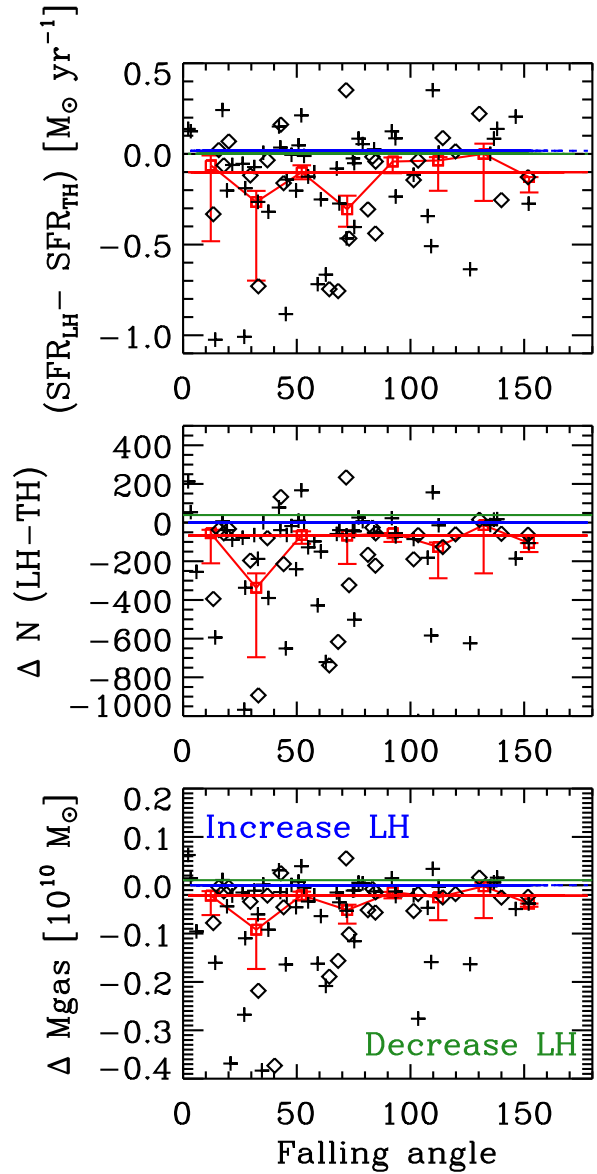}
\caption{Total integrated $\rm SFR$ and number of particle (top and bottom, respectively) differences between the gas particles of the leading and trailing as a function of the falling angle. 
The blue lines show the median from using the $\rm 20\%$ of the particles in each galaxy that are furthest into the leading and trailing halves.
The red line shows the median when all particles in each half are used.
Red squares show the median at each bin, the errors are the median divided by the number of galaxy at each bin.
Diamonds show galaxies residing in relaxed clusters, while crosses in less relaxed clusters (see section \ref{sec_dynamical_state}). The solid green line shows the median of the control sample.}
\label{fig_percentiles}
\end{figure}

\subsubsection{Integrated differences}
\label{sec_percentages}
In this subsection, we present the comparison of the integrated properties considering the same number of particles on both sides. 
We select a fixed percentage of 20$\%$ of the total number of gas particles per side, choosing the most distant ones with respect to the center of mass, hereafter referred as quintiles.

On either side this selection reduces the effects of intrinsic asymmetries between the two halves, such as different number of particles, gas mass, etc., which could dominate the overall differences seen in Fig. \ref{fig_main}.
In the upper, middle, and lower panel of Fig. \ref{fig_percentiles}, each individual symbol shows the difference of the integrated $\rm SFR$, number of particles, and gas mass over all the particles located in the leading and trailing half, respectively. 
In each panel of Figure \ref{fig_percentiles}, the solid green line shows the median value of the control sample.
The red line in the middle panel shows the median of the difference between the number of particles residing in each half.
Most of the satellites ($\rm 78\%$) show an overabundance of gas particles in the trailing half with respect to the leading half. Only $\rm 22\%$ of them show the opposite enhancement.

The red line in the top panel shows the median $\rm SFR$ difference if all gas particles are considered in each half, while the blue line indicates the mean difference from the quintiles.
The $\rm SFR$ difference increases from  $\rm -0.10^{+0.02}_{-0.06}$ (red line) to $\rm +0.017^{+0.009}_{-0.003}$ (blue line) showing that the difference in mass or number of particles between halves can mask the $\rm SFR$ difference.
The median value of the quintiles is positive and close to $\rm 2\%$, with an error three times higher for the positive than negative values. 

These results suggest that positive $\rm SFR$ differences between leading and trailing halves would be measured in observations if the extreme galaxy sides, of equal area, are analyzed, since differences are stronger.
A similar median of the $\rm SFR$ enhancement is measured if different percentages of the number of particles are considered.
The median $\rm SFR$ of the leading half (20$\%$ of particles) is 2$\%$ higher than the $\rm SFR$ overall the galaxy. To measure this difference might be challenging in observational works as discussed in section \S \ref{sec_discusion}.

In the bottom panel, the mass difference between the gas particles of the trailing and leading half is shown. Although each gas particle has a different mass, the trend of the gas mass difference is pretty similar to the one of the number of particle difference. Here, all the gas particles found by the $\rm SUBFIND$ algorithm with $SFR > 0$ are considered (see section \S \ref{ssec_satellite}).

Following the previous results, in Figure \ref{fig_percentiles} the galaxies located within one virial radius of the group or cluster center are shown.
If all galaxies up to 2 (3) virial radius are taken into account, the number of selected galaxies triples and the fraction of galaxies with a more abundant leading half rises from $\rm 22 \%$ to $\rm 34\, (38) \%$, while the median $\rm SFR$ difference of the quintiles is close to zero, $\rm +0.005^{+0.004}_{-0.002}$ ($\rm +0.004^{+0.003}_{-0.002}$) and negligible in observations (below 0.5$\%$ of the total galaxy $\rm SFR$). 
This suggests that the RP effect, which is causing the overabundance in the trailing half, is less effective in the outer cluster regions (above one virial radius).

\subsubsection{Gas phase metallicities}

For the galaxy sample described in section \S \ref{sec_sample}, located within one virial radius, the median metallicity and oxygen abundance enhancement are $\rm \langle \delta Z^m_E \rangle = 3\%^{+1}_{-1}$, $\rm \langle \delta O/H^m_E \rangle = 3\%^{+1}_{-1}$, respectively.
The median of the mass-weighted average metallicity enhancement $\rm \langle \delta Z^w_E \rangle = 4\%^{+1}_{-1}$, and Oxygen abundance enhancement are
$\rm \langle \delta O/H^w_E \rangle = 4\%^{+1}_{-1} $, respectively.
There is a marginal difference of the chemical enrichment, of $\rm 3-4\%$, between the leading and trailing halves with respect to the overall galaxy value. 
The gas of the leading half is richer $\rm 3-4\%$, in terms of metals and Oxygen abundance, than the trailing half.
There are two cases with falling angles around $\rm 50$ degrees,
in which the median metallicity or oxygen abundance enhancement is higher than $\rm 50\%$ of the median metallicity or oxygen abundance of the complete galaxy. The same satellites show the highest $\rm SFR$ differences and enhancements (see figure \ref{fig_jellyfishEAGLE}).\\

These differences in $\rm SFR$, $\rm SFE$ and pressure do not translate into significant differences in the stellar populations properties (such as stellar ages and metal abundances) as the amount of new star formation is on average negligible compared to the existing stars.

\section{Connection with observable quantities}

In this section we discuss the feasibility of confronting the results presented in section \S \ref{sec_results} with observations. 
The first subsection studies a case that might be applicable to observations. The second and third sections are dedicated to the dependency of the $\rm SFR$ enhancement with cluster and intrinsic properties of the satellites, while the fourth one compares our results with main-sequence galaxies of the EAGLE simulation.

\subsection{Maximum anisotropy cut}
\label{sec_perrito}

Firstly, we need to find the appropriate plane to divide the halves that is feasible in observations. 
Hereafter, we use the three-dimensional positions and velocities,
while projections effects and analysis of mock images will be presented in a separate article.
Future photometric surveys of broad, medium and narrow bands will be able to trace the three-dimensional structure of thousands of groups and clusters. An example is the J-PAS survey reaching a precision of $0.003\times(1+z)$ \citep{ascaso16}.

As analyzed in \cite{troncoso16b}, 
the most simple case is choosing the plane perpendicular to the 
radial vector of the center of potential with respect to its central galaxy (see middle panel of Figure \ref{fig1_scheme}). 
Yet, when analyzing the $\rm SFR$ differences using this plane, we find no significant difference between the two halves or neither a correlation with the velocity case. In some cases, we even find an enhancement in the half pointing away from the central galaxy. 

In order to look for another possible observationally applicable plane to divide a galaxy in two halves, we perform a  search in all the possible orientations, homogeneously sampling the sphere, for the plane that maximizes the mass-weighted $\rm SFR$ difference or enhancement between the trailing and leading half, Eq. \ref{mw_sfre}. Hereafter, this maximum difference is dubbed $\rm SFR_{E}^{max}$.

\begin{figure}
\includegraphics[width=1.05\columnwidth, trim=1.5cm 1cm 0cm 0cm]{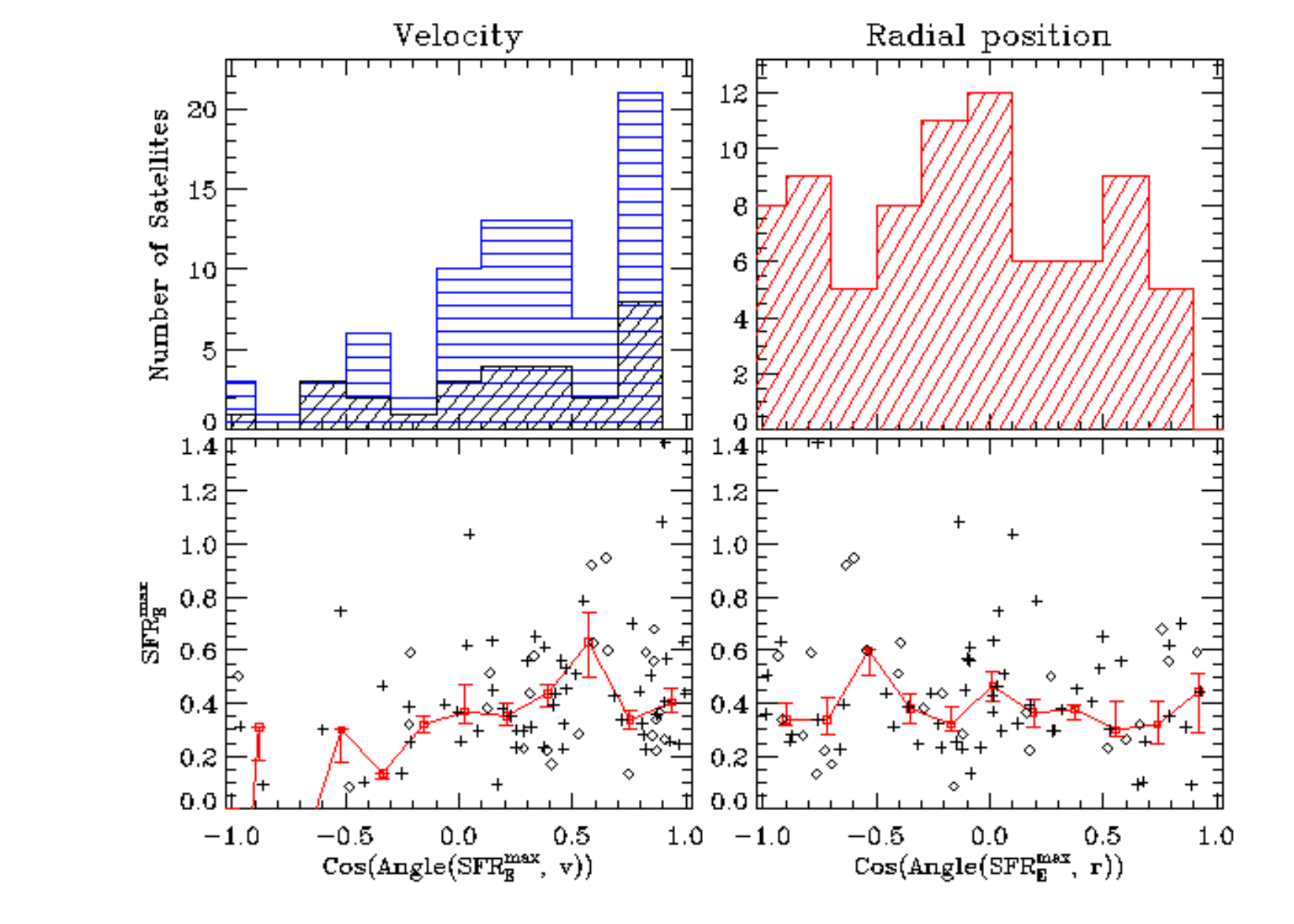}
\caption{Direction that maximizes the mass-weighted $\rm SFR$ enhancement.
The upper-left (right) panel shows the histogram of the angle $\alpha$, i.e. the angle between the vector perpendicular to the plane that maximizes the mass-weighted $\rm SFR$ difference between the two halves and the velocity (radial) vector, both with respect to the central galaxy. Bottom panels show the maximum $\rm SFR$ enhancement as a function of the alignment angle $\alpha$, with respect to the velocity (left) or radial position (right).
Diamonds show galaxies residing in relaxed clusters, while crosses in less relaxed clusters (see section \ref{sec_dynamical_state}).}
\label{fig_perrito}
\end{figure}

In Figure \ref{fig_perrito}, we focus the attention on the galaxy sample described in section \S \ref{ssec_satellite}, and satellites located within one virial radius.
The left upper panel of Figure \ref{fig_perrito} shows the histogram of the alignment angle $\alpha$, defined as the angle between the vector normal to the plane that maximizes the mass-weighted $\rm SFR$ difference and the velocity vector (see right panel of Figure \ref{fig1_scheme}).
In the case of perfect alignment, alignment angle equal to zero, the vector normal to the plane that maximizes the difference corresponds exactly to the direction of the velocity with respect to the ICM. 
The blue distribution is broad and peak around an angle of zero degrees.
 It suggest that statistically the plane 
that maximizes the difference indicates the velocity with respect to the ICM, but it cannot used in a case-by-case form.

The right upper panel of Figure \ref{fig_perrito} shows the analogous angle for the radial vector. It shows that the distribution is nearly random suggesting that there is little or non alignment between the plane that maximizes the enhancement and the radial vector.
This result suggests that even if we could measure the tri-dimensional structure of the cluster, the RP effects cannot be measured without knowing the velocity of each satellite mainly because there is no alignment between the position and velocity direction with respect to the central galaxy. 
The bottom panels of Figure \ref{fig_perrito} show the maximum $\rm SFR$ asymmetry as a function of the alignment angle in each case, with respect to the velocity or position vector.
In the case of the position vector ({\it right panel}), no trend is observed, while a weak one exits between small velocity alignment angles and higher values of maximum $\rm SFR$ asymmetry {\it left panel}. 
It shows that higher asymmetry is likely to be caused by the galaxy velocity, i.e. RP.

By splitting our sample in four bins of falling angle in Figure \ref{fig_perrito},
we found that the satellites with a falling angle below 45 degrees tend to show an alignment ($cos \,\alpha >0.75$), while the other bins show no preferential alignment.
The black histogram shows this sub sample.
The same group of galaxies show the highest $\rm SFR$ enhancements (see Figure \ref{fig_main} and \ref{fig_jellyfishEAGLE}).
No dependency with the relaxation index is observed.

We visually inspected ten satellites of stellar masses above $10^{10}[M_{\odot}]$ showing the highest asymmetry $\delta  SFR^{max}_{E} > 0.6$, namely ID 11529078, 9440185, 16300531, 10751313, 6082966, 14895218, 11525815, 15691064, 4561773, 10687442.
They are classified as jellyfish galaxies according to the criterion described in section \ref{sec_jellyfish}.
In the case of ID 16300531 and 14895218, the observed asymmetry is mild with respect to the other eight examples.
We visually inspected the satellites of stellar masses above $10^{10}[M_{\odot}]$ because the optical mock images are available in the EAGLE database and the number of gas particles is high enough to perform a visual inspection with high statistics.

We study the correlation between the mass-weighted $\rm SFR$ enhancement ($\rm \delta  SFR^{mw}_{E}$) and the maximum mass-weighted $\rm SFR$ enhancement ($\rm SFR^{max}_{E}$).
We measure a moderate positive correlation between the $\rm SFR^{max}_{E}$ and the positive values of $\rm \delta  SFR^{mw}_{E}$ or $\rm \delta  SFR^{m}_{E}$, with a Pearson correlation coefficient, $\rm  R=0.48, R=0.46$, respectively.

No correlation is observed and measured for the logarithmic ratio excess of the same quantities ($\rm SFR^{mw}_{LRE}$ and $\rm SFR^{max}_{LRE}$).

These results suggest that an observational analysis using the plane that maximizes the $\rm SFR$ difference between the two galaxy halves would provide an indication of the direction of the peculiar velocity of the galaxy, and would help to reconstruct the effect of RP.
In forthcoming article by \citet{silvio}, we have applied a this method, projected in the sky plane, to the SDSS photometric data \citep{sdss09}.

\subsection{Dependence with cluster properties} \label{sec_dep_cluster}

We now consider the selected satellite sample of section \S \ref{sec_sample} located within 3 virial radii (211 galaxies), i.e. satellites with more than one hundred particles in their stellar and gas components, stellar masses above $\rm 10^9 M_{\sun}$, residing in groups more massive than $ \rm 10^{13.6} M_{\sun}$.
It corresponds to 95$\%$ of the total satellites eligible for this analysis (223), the ones that are excluded are located above three virial radii.

Here we study whether the mass-weighted $\rm SFR$ enhancement depends on the falling angle, group cluster mass, the distance of the satellite to the center of potential of the group/cluster, the relaxation index, and the alignment. The last one is quantified with the angle $\alpha$, angle between the velocity vector and normal vector to the plane that maximizes the mass-weighted $\rm SFR$ enhancement. This angle is schematized in the right panel of Figure \ref{fig1_scheme} and the histogram of the alignment is showed in the upper-left panel of Figure \ref{fig_perrito}.

\begin{figure}
\includegraphics[width=\columnwidth, trim=4.5cm 14cm 12.5cm 3.5cm]{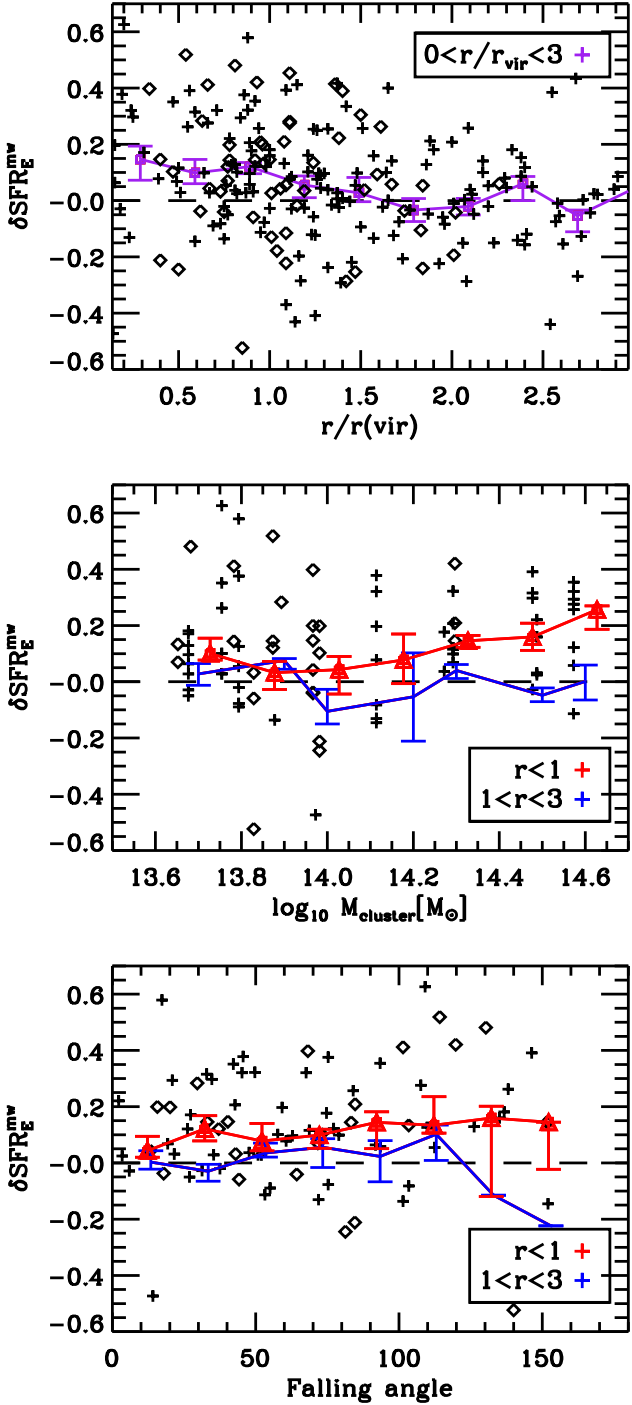}
\caption{Mass-weighted $\rm SFR$ enhancement as function of cluster properties, measured in the first case, namely {\it velocity cut} (see text).
{\it Top}, {\it middle}, and {\it lower} panel shows the median $\rm SFR$ enhancement as a function of the radial distance to the central galaxy normalized to the virial radius, cluster mass, and falling angle, respectively. 
Diamonds show galaxies residing within one virial radius of relaxed clusters, while crosses in less relaxed clusters (see section \ref{sec_dynamical_state}).
Red diamonds and blue crosses indicate the median of the $\rm SFR$ enhancement for satellites within one and three virial radii, respectively.
Purple squares indicate the median of the $\rm SFR$ enhancement for satellites within three virial radii. 
The error bars show the one sigma error.} \label{fig_pp_cluster}
\end{figure}

Figure \ref{fig_pp_cluster} shows the dependence of the mass-weighted $\rm SFR$ enhancement, $\rm SFR_E^{mw}$, as a function of the radial distance to the central galaxy (top panel), cluster mass (middle panel), and falling angle (bottom panel). 
Red and blue lines show the median $\rm SFR_E^{mw}$ of the galaxies located within 1,  and, 1 and 3 virial radii, respectively. The first sample is composed of 79 galaxies, while the second one of 132. 
The error bars indicate the one sigma error on the median.
In the top panel, the purple line indicates the median $\rm SFR_E^{mw}$ of the galaxies located within 3 virial radius. It  appears to decrease with increasing radius up to $r=1 \times r_{vir}$, above this radii the trend is consistent with zero. 
In the middle panel the satellites residing within one virial radius show an increase of the $\rm \delta SFR_E^{mw}$ with cluster mass, while for ones located within 1 and 3 virial radii the trend vanishes.
In the bottom panel, the $\rm \delta SFR_E^{mw}$ tends to keep constant with falling angles, for both group of satellites, located within one or  one and three virial radii. The first group presents a higher median around 0.1, while the second one around zero. 
The satellites located further out and with falling angles above 110 degrees show negative $\rm \delta SFR_E^{mw}$, i.e. its trailing half is more star-forming than the leading half. 

A similar trend for the logaritmic ratio excess of the mass-weighted $\rm SFR$ ($\rm SFR_{LRE}^{mw}$) as a function of the virial radii, cluster mass and falling angle is observed.

Regarding the alignment, the angle between the velocity vector and the normal vector to the plane that maximizes the $\rm SFR$ difference, we observe no dependency with cluster mass or relaxation index.
  
There is small signal of the $\rm SFR$ enhancement to correlate with the falling angle for galaxies at $\rm r<3\,r_{vir}$, however a larger sample of simulated cluster galaxies is required to conclusively claim that.

\subsection{Comparison with other intrinsic galaxy properties} \label{sec_dep_gp}

We also study whether the asymmetry in $\rm SFR$ is also visible in other galaxy properties. 
In this section we mention the overall differences found, but relegate the 
figure to the Appendix (Fig.\ref{fig_sfrenh_gp}).  We studied the mass-weighted $\rm SFR$ enhancement or percentage excess as a function of the galaxy intrinsic properties, $\rm SFR$, stellar mass, specific star formation rate ($\rm sSFR$), star-forming gas mass, and total gas mass. 
A positive median excess, with respect to the control sample, is observed on all properties for satellites located within one virial radius, except by the negative enhancement observed in the lowest values of each quantity.

In the $\rm SFR$ case, it might be a result of  reaching the resolution limit of the simulation.
In the $\rm sSFR$ case, it is remarkable that $\rm RP$ also affects the gas of less active satellites.
Another possible explanation is such small satellites ($\rm M_*< 10^{9.5} M_{\sun}$) located within one virial radius, are not longer forming stars because their gas was fully consume or lost by RP and other process related to galaxy assembly itself like mergers, starvation, etc.
\\
It is worth to mention that $\rm 60\%$ of our sample has stellar masses within $\rm M_*< 10^{9.5-10.5} M_{\sun}$, which is dubbed the intermediate mass sample.
Outside this range the statistics are poorer.

\subsection{Comparison with EAGLE main sequence galaxies} \label{sec_ms}

As explained in section \S \ref{sec_controlsample}, we select the main sequence galaxies with $\rm sSFR> 0.01 \,Gyr^{-1}$ \citep{furlong15} and stellar masses above $\rm 10^9 M_{\odot}$, and excluding satellites ($SubGroupNumber \,= 0$). 

\begin{figure}
\includegraphics[width=\columnwidth, trim=3.5cm 13.5cm 12cm 3cm]{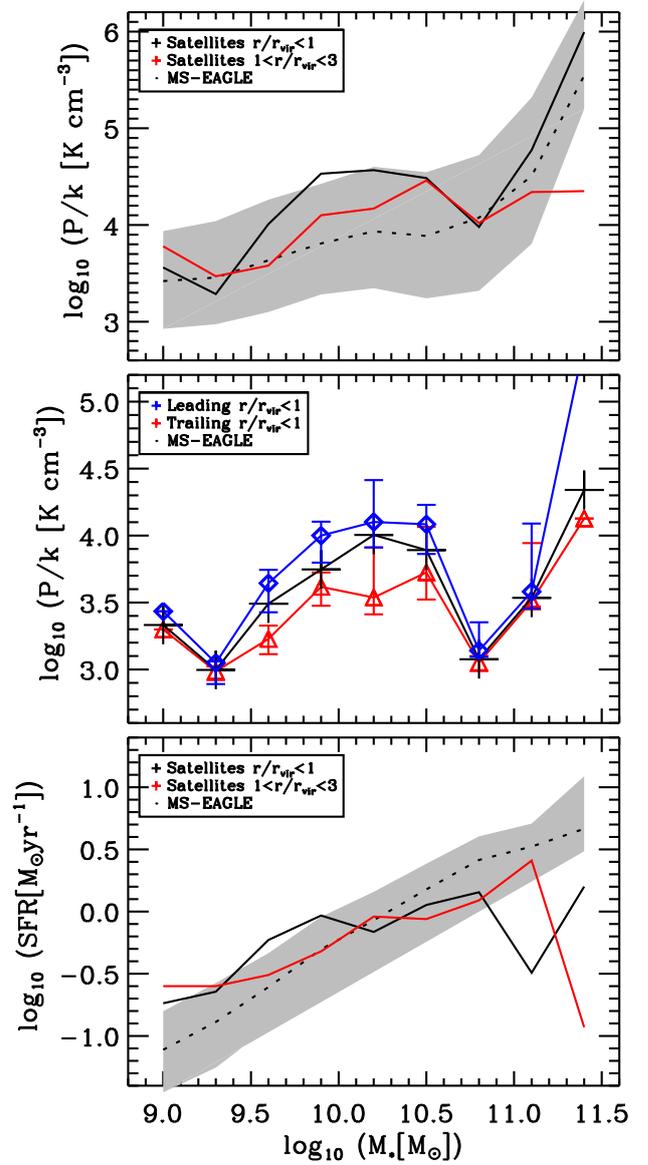}
\caption{Comparison of the pressure and $\rm SFR$ between our sample of cluster galaxies described in section \S \ref{sec_sample} and $z=0$ main sequence galaxies in the EAGLE simulation, according to the definition of \citet{furlong15}, $\rm sSFR > 0.01\,  Gyr^{-1}$ and excluding satellites.
{\it Top panel:} $\rm SFR$-weighted average pressure as a function of stellar mass for main sequence (dotted line) and our cluster galaxies in EAGLE located within one (black line) and three virial radii (red line). The shaded region shows the dispersion of the data.
{\it Middle panel:} the median pressure of the galaxy (black crosses), leading (blue diamonds), and trailing halves (red triangles) is plotted versus the galaxy stellar mass for our satellites located within one virial radius. Errors bars show the error on the median.
{\it Bottom panel:} location of the cluster galaxies in the $\rm log_{10} SFR- log_{10}M_*$ plane.
Black and red lines show the median of the cluster galaxies located within one and three virial radius, respectively. The dotted line shows the $\rm SFR$ of the main-sequence galaxies of the EAGLE simulation, while the gray shaded regions marks the dispersion of the data.}
\label{fig_ms}
\end{figure}

The top panel of Figure \ref{fig_ms} shows the $\rm SFR$-weighted average pressure as a function of stellar mass for main sequence (dotted line) and cluster galaxies in EAGLE located within one (black line) and three virial radii (red line), selected according the criteria defined in section \S \ref{sec_sample}.
The shaded region shows the dispersion of the main sequence galaxies in EAGLE. 
The dispersion of the satellites, located within one and three virial radii, is similar to the one shown in the gray shaded region. 

In the middle panel of Figure \ref{fig_ms}, the median pressure of the satellite galaxies (black crosses), and for the leading (blue diamonds), and trailing halves (red triangles) are plotted as a function of the stellar mass.
The errorbars show the error on the median.
The median of the leading halves is higher than the trailing halves and also higher than the galaxy as a whole for all the stellar mass bins.
Only satellites located within one virial radius are considered.
A similar trend, with smaller differences between the samples, is observed when all satellites are considered.
Satellites in the mass range $\rm 10^{9.5-10.5} \rm M_{\sun}$ show the strongest differences.
The lower limit $\rm 10^{9.5} \rm M_{\sun}$ might be indicating that we are reaching the limit of the simulation to trace the RP stripping, indeed these galaxies are the ones described with the lowest number of gas particles in the simulation $\rm 242^{+32}_{-28} $ with respect to the median of the full satellite sample $\rm 519^{+81}_{-39}$.
The upper limit $\rm 10^{10.5} \rm M_{\sun}$ indicates that satellites with stellar masses above it, might constitute a different population of particular properties.
\cite{wright19} show that the quenching timescale of satellites of masses within the range $\rm 10^{9.5-10.5} \rm M_{\sun}$ is larger than for satellites outside this range. It suggests that the quenching timescale of galaxies outside this stellar mass range is so short that we are not able to trace the RP stripping.
 
Other explanation is that these satellites are displaced of the main sequence, as it is discussed in the following.
The bottom panel indicates the position of the main sequence galaxies in the logarithmic $\rm SFR$, stellar mass plane.
Red, black lines show the median of the cluster galaxies located within one and three virial radii, respectively.
The dotted line shows the $\rm SFR$ of the main-sequence galaxies of the EAGLE simulation, while the gray shaded region marks the dispersion of the data.
The dispersion of the satellites, located within one and three virial radii, is similar to the one shown in the shaded area. Except for the most massive galaxies ($\rm > 10^{11} \rm M_{\sun}$), where the dispersion is higher due to few statistics. 
Our satellites show higher $\rm SFR$ at low-masses, most likely due to the imposed limit on the number of particles ($\rm N_{gas} > 100$), while higher masses lie below the median.
This shows that the satellites with $\rm SFR$ lower than the main sequence galaxies also show pressure enhancement, confirming that RP is increasing the ISM pressure even though there is $\rm SFR$ suppression.

Another aspect to consider is that in the case of satellites of masses above/below $\rm > 10^{10.5}/10^{9.5} \rm M_{\sun}$ we might not be able to measure RP because we are reaching the limits of the simulation in two senses i) we sample few galaxies or ii) only the ones with low $\rm SFR$, hence few gas particles are available to trace the RP.

\section{Discussion} \label{sec_discusion}

We have measured the differences of pressure, $\rm SFR$, metallicity, oxygen abundance and age between the leading and trailing halves for the set of EAGLE satellites defined in section \S \ref{sec_sample}, considering different intervals of cluster mass, falling angles, and radial distance to the central galaxy.

The left, middle and right panels of Figure \ref{fig_main} show that the leading half presents an enhancement of the median $\rm SFR$, gas mass-weighted $\rm SFE$, and median pressure with respect to the trailing half. 
The middle panel of Figure \ref{fig_percentiles} puts in evidence that most of the analyzed satellites present and overabundance of gas particles in the trailing half with respect to the leading half. 
This suggests a transport of gas from the leading to the trailing half due to the compression of the ICM.
In absolute terms, the top panel of Figure \ref{fig_ms} shows that, overall, the pressure of the analyzed satellites tend to lie above the median values of the main-sequence galaxies. Yet, the satellite and main sequence pressures are consistent with the same distribution.
The middle panel of the same figure shows an excess of the median pressure of the leading (blue diamonds) with respect to the overall galaxy (black crosses) which in turn is higher than the trailing half (red triangles).
This indicates that the pressure enhancement is driven by the leading half (blue crosses), most likely due to compression of the ISM by the interaction with the ICM.

Depending on the chosen way to measure the $\rm SFR$ excess or enhancement in the simulation, we find different dependence of the effect as a function of the falling angle. 
For example,  the bottom panel of Figure \ref{fig_pp_cluster} shows a nearly constant enhancement for galaxies within one virial radius, independently of the falling angle. 
The left panel of Figure \ref{fig_main} shows that galaxies falling with angles above 90 degrees present the highest median $\rm SFR$ excess.
The former result reflects the global value of each half, while the latter one can be used to select jellyfish type galaxies from the simulation, whose mock images are similar to those typically seen in observational works of jelly-fish galaxies \citep{pattarakijwanich16,poggianti16}.
We can select the most disrupted galaxies by ranking all cluster galaxies with its median percentage excesses ($\rm SFR_E^{max}$). 
These RP affected satellites contribute to the scatter observed in the scaling relations of local galaxies residing in groups or clusters \citep{lin14,koyama13,peng10}.

\citet{yun19} analyzed the satellites of  massive groups, $\rm M\, [M_{\sun}]>10^{13-14.6}$ in the ILLUSTRISTNG100 simulation.
In this work their findings are similar to ours 

in terms of the dependence of the abundance of jellyfish galaxies as a function of stellar mass and separation from the cluster centre.
The median pressure of the leading half (middle panel of Fig. \ref{fig_ms}) tends to be significantly higher for galaxies in the stellar mass range $\rm 10^{9.5-10.5} [M_{\sun}]$ with respect to the main sequence galaxies in  EAGLE, and with respect to their own trailing half.
The $\rm SFR$ enhancement reaches its maximum in the range $\rm 0.2<r/r_{vir} < 1$.
These ranges are in agreement with the ones presented in \citet{yun19} who find that the abundance of jellyfish galaxies peaks at low satellite masses above $\rm 10^{9.5} [M_{\sun}]$ and at $\rm r/R_{200}>0.25$.
The lower resolution limit of the galaxy sample used in \citet{yun19} is $\rm 10^{9.5} [M_{\sun}]$.

\subsection{Feasibility of the measurements with Observations}
\label{disc_perrito}

Figure \ref{fig_perrito} shows that vector normal to the plane that maximizes the $\rm SFR$ differences, in most cases, is well aligned to the velocity vector.
Hence, using this plane and the maximum $\rm SFR$ difference it is possible to select the galaxies that are most extremely asymmetric due to the effect of RP.
Hence, the plane that maximizes the $\rm SFR$ difference is recommended for statistical and observational studies of RP effects.

The median or the mass-weighted $\rm SFR$ enhancements are preferentially positive, indicating that the leading half is more star-forming than the trailing half,
for satellites that satisfy the following, located within one virial radius, with overall mass-weighted $\rm SFR$  greater than $\rm 0.5\, M_{\odot}/yr$, a stellar mass  greater than $\rm M_* > 10^{9.5} M_{\odot}$, and  gas masses greater than $\rm M_{gas} > 10^{9} M_{\odot}$. Galaxies within one virial radius show higher $\rm SFR$ enhancement than galaxies located outside this radius. 
The $\rm SFR$ enhancement is higher in galaxies residing in massive clusters and within one virial radius. This dependence vanishes when all satellites are considered. 
No trend is observed as a function of the falling angle.
Hence, the galaxies with asymmetric star-formation are most likely to be found with properties within the limits mentioned above.

The moderate correlation between $\rm SFR_E^{max}$ and $\rm \delta SFR_E^{mw}$, or $\rm \delta SFR_E^{m}$,
indicates that an observational analysis that searches for the maximum $\rm SFR_{E}^{max}$ difference and selects the quintile with the highest $\rm SFR_{E}^{max}$, would help to find the galaxies most affected by RP without visually inspecting all cluster members.
Furthermore, we have visually inspected, according the criteria defined in section \S \ref{sec_jellyfish}, the galaxies selected with the highest asymmetry ($\rm SFR_{E}^{max} > 0.6 $) and stellar mass above $10^{10}\, M_{\odot}$, classifying them as jellyfish galaxies.

Figure \ref{fig_sfrmax_gp} shows the maximum mass-weighted $\rm SFR$ difference as function of intrinsic galaxy properties and of cluster properties. 
Red and blue lines show the median of galaxies located within one and three virial radii, respectively. 
In each panel, the solid green line shows the median value of the control sample.
In every panel, the median of the satellites located within one virial radius is higher than the control sample, except by $\rm log_{10} sSFR [Gyr^{-1}]> -1$.
For satellites located within one virial radius, a positive correlation between the $\rm SFR_{E}^{max}$ and the stellar mass is observed, while a negative one is seen for the $\rm sSFR$, and radial distance to the central galaxy. 
Although we found a moderate correlation between $\rm SFR_{E}^{mw}$ and $\rm SFR_{E}^{max}$, no correlation is observed between the $\rm SFR_{E}^{max}$ and the cluster mass or falling angle.
If all satellites located up to three virial radii, then these trends vanish.
No dependence with other galaxy properties such as $\rm SFR$, gas mass is observed.
\\
Simultaneous $\rm SFR$ and gas maps of cluster galaxies, and particularly jellyfish galaxies would allow to study a correlation between the $\rm SFR$ and $\rm SFE$ enhancements. For example, combining MUSE with ALMA maps of jellyfish galaxies would allow such measurements and confirmation or refutal of our predictions (see Figure \ref{fig_main}).

\subsection{Properties in the Local Universe}

The mass-weighted $\rm SFR$ enhancement is the percentage difference between the leading and trailing half with respect to the overall value, see Eq. \ref{eq_sfr}.
The $\rm SFR_{E}^{mw}$ values reported in Table \ref{tab_1}, of around 10$\%$, indicate that our mass-weighted $\rm SFR$ individual measurements of each half must be more precise than this, or otherwise the RP effects cannot be detected in observational studies.

Using photometry it could be possible to use blue band magnitudes to trace the $\rm SFR$. 
IFU observations allow to find asymmetries in $\rm SFR$ in individual galaxies.
However, this would be quite challenging because in order to achieve accurate $\rm SFR$ measurements it is necessary to detect the $\rm H_\beta$ line with a reasonable signal to noise, which is typically obtained by integrating three times longer than for $H_\alpha$, depending on the extinction.
Since in this work we are proposing to integrate the $\rm SFR$ in galaxy halves,  not using the individual spaxels, the signal increases as $\rm \sqrt{N_{spx}/2}$, where $\rm N_{spx}$ is the number of total spaxels of the IFU.

One way to loosely mimic this type of observations is achieved by selecting only the particles above a certain $\rm SFR$ threshold.  This makes the integrated $\rm SFR$ of the leading half  20$\%$ higher than the trailing one, which would be even easier to detect in IFU observations.

\subsection{Misleading results using the most simplistic radial case.} \label{sec_badobs}

 \cite{troncoso16b} reported the results using a simplistic \texttt{radial cut} proposed in section \ref{sec_mns} using the EAGLE satellites and the same selection reported in section \S \ref{sec_sample}. 
Figure 2 of their work shows that the behavior of the $\rm SFR$ differences as a function of the falling angle of the \texttt{velocity} and \texttt{radial cut} are clearly dissimilar.
This suggests that  observations could show the opposite results  due to projection effects on the plane used to cut the galaxy in halves. 
In observations, the plane perpendicular to the radial position is the most simplistic one that can be used because the three-dimensional velocity vector with respect to its central galaxy is unknown. Hence, the results of the observations might be biased in this case. \citet{yun19} found a similar result, by analyzing the correlation between the gas tails of the TNG jellyfish galaxies (2600) and the satellite bulk velocity and position with respect to the host center (see Fig. 7). They also found no alignment in the case of the position, while for the velocity there is a clear indication to form angles around 180 degrees.

\section{Conclusions} \label{sec_conclusions}

Using the biggest simulation of the EAGLE consortium, we measure differences in the physical properties of halves of galaxies that are falling into clusters (see Figure \ref{fig_pressure}).
We dissect the galaxies that are star-forming in two halves by using three planes that define our three cases of study: \texttt{velocity cut}, \texttt{radial cut}, and \texttt{maximum anisotropy cut} (see Figure \ref{fig1_scheme}). The first plane divides the satellites using the plane perpendicular to the three-dimensional velocity vector, the second one uses the plane perpendicular to the three-dimensional distance to its central galaxy, while the third one finds the plane that maximizes the $\rm SFR$ difference between both halves.

In the first case, using the \texttt{velocity cut}, we observe an enhancement of the $\rm SFR$s, $\rm SFE$s, and pressure of gas particles in the leading half with respect to the trailing one (see Figure \ref{fig_main}), while the number of gas particles in the trailing half is systematically higher than in the leading one (see Figure \ref{fig_percentiles}).

We suggest that the measured differences are evidence of RP acting on the leading half, transporting the gas from the leading to the trailing half, enhancing the ISM pressure in the leading half, and consequently boosting its $\rm SFR$.
This effect depends on the cluster properties as well as galaxy intrinsic properties as follows.

\begin{itemize}
\item For satellites located within one virial radius, we observe a positive correlation between the $\rm SFR$ enhancement and the stellar mass, the cluster mass, the gas mass, and the $\rm SFR$.  
We find a negative correlation for the specific star-formation rate and radial distance to the central galaxy.
If all satellites, including those located farther than the virial radius, are considered then the trends vanish (see Figure \ref{fig_pp_cluster} and Figure \ref{fig_sfrenh_gp}).

\item These differences are small, around $10\%$ of the overall mass-weighted $\rm SFR$ value, hence it might be difficult to detect them using data from ground based large scale surveys (see Figure \ref{fig_main}).
In the case of photometric surveys, this effect could be measured using blue bands as a $\rm SFR$ tracer, while for IFUs the situation improves when halves of the galaxy and not individual spaxels are analyzed.

\item When dividing the sample according to the dynamical state of the host cluster, we do not find a difference of the median or mass-weighted average $\rm SFR$ enhancement in relaxed compared to less-relaxed clusters. There are more infalling galaxies and of small falling angles in less-relaxed than in relaxed clusters (see Figure \ref{fig_main}). 
Both types of clusters show similar numbers of galaxies receding away from the cluster centre. 

\item By comparing our sample of satellites with the main-sequence galaxies of the EAGLE simulation, we found that of all the satellites selected here have a higher pressure, which is even higher in the leading half compared to the trailing one. 
We suggest that it is the compression of the ISM due to the interaction with the ICM that drives the pressure enhancement.
Our sample of satellites is overall more star-forming than main-sequence galaxies up to $\rm 10^{10.5} \, [M_{\sun}]$, above this limit our satellites are suppressed in SF but still show ISM compression in their leading half (see Figure \ref{fig_ms}).
These RP affected satellites contribute to the scatter observed in the global scaling relations of local galaxies residing in groups and clusters \citep{lin14,koyama13,peng10}.
\end{itemize}

The \texttt{radial cut} proposed clearly fails to detect the real physics occurring behind the RP effects. 

\texttt{The maximum anisotropy cut}, that uses the plane that maximizes the $\rm SFR$ difference between the two halves, allows us to
find that the vector normal to this plane is aligned with the three-dimensional velocity of the galaxy (see Figure \ref{fig_perrito}).
This finding suggests this is the most suitable way in observations to detect and study RP effects on statistical samples of observed galaxies.

This alignment is also reflected in a moderate correlation between the maximum $\rm SFR$ difference and the $\rm SFR$ difference found using the velocity direction of the satellite to halve the galaxy.

Selecting the galaxies with the highest $\rm SFR$ difference, 
would allow to find the galaxies most affected by RP without visually inspecting all cluster members.
This automatic method can be used to select asymmetric galaxies, without visually inspecting all satellites, in the era of large scale surveys (J-PAS, BOSS, LSST, etc.) and future ones.

\section*{Acknowledgements}
We thank the anonymous referee for her/his useful comments and timely responses.
PTI acknowledges financial support from FONDECYT POSTDOCTORAL 3140542, DIUA 162-2019 of Universidad Aut\'onoma de Chile, and ANILLO-ACT-1417. 
PTI further acknowledges her better half, Zoila $\&$ Juan for taking care of her baby Logan while she wrote this article and Pontificia Universidad Cat\'olica for hosting her during the creation of this work.  
CL is funded by an ASTRO 3D Senior Fellowship via the Australian Research Council Centre of Excellence CE170100013.
NP \& SC acknowledge support from a STFC/Newton Fund award (ST/M007995/1 - DPI20140114).
SC acknowledges the support of the ``Juan de la Cierva formacion'' fellowship (FJCI-2017-33816).
We acknowledge the Virgo Consortium for making their simulation data available. The EAGLE simulations were performed using the DiRAC-2 facility at Durham, managed by the ICC, and the PRACE facility Curie based in France at TGCC, CEA, Bruy\'eresle-Ch\^atel.
The Geryon cluster at the Centro de Astro-Ingenieria UC was extensively used for the calculations performed in this paper. 
BASAL CATA PFB-06, the Anillo ACT-86, FONDEQUIP AIC-57, and QUIMAL 130008 provided funding for several improvements to the Geryon cluster.




\begin{thebibliography}{99}


\bibitem[\protect\citeauthoryear{Abazajian et al.}{2009}]{sdss09}
Abazajian K. N., et al. The Seventh Data Release of the Sloan Digital Sky Survey.
{\em ApJS}, {\bf 2009}, {\em 182}, 543

\bibitem[\protect\citeauthoryear{Aghamousa et al.}{2016}]{desi16}
DESI Collaboration: The DESI Experiment Part I: Science, Targeting, and Survey Design. Aghamousa, A. et al. 2016, arXiv:1611.00036

\bibitem[\protect\citeauthoryear{Ascaso et al.}{2016}]{ascaso16}
 Ascaso, B. et al. An Accurate Cluster Selection Function for the J-PAS Narrow-Band wide-field survey. {\em MNRAS} {\bf 2016}, {\em 456}, 4291-4304.
 
\bibitem[\protect\citeauthoryear{Baes et al.}{2011}]{baes11}
Baes, M. et al. Efficient Three-dimensional NLTE Dust Radiative Transfer with SKIRT. {\em MNRAS} {\bf 2015}, {\em 450}, 1937-1961.

\bibitem[\protect\citeauthoryear{Baldry et al.}{2010}]{baldry10}
Baldry, I. K. et al 2010.
Galaxy And Mass Assembly (GAMA): the input catalogue and star-galaxy separation.
{\em MNRAS} {\bf 2010}, {\em 404}, 86

\bibitem[\protect\citeauthoryear{Banerji et al.}{2010}]{banerji10} Banerji, M. et al. 2010.
Galaxy Zoo: reproducing galaxy morphologies via machine learning {\em MNRAS} {\bf 406}, {\em 342}, 
  
\bibitem[\protect\citeauthoryear{Bellhouse et al.}{2019}]{bellhouse19}
Bellhouse, C. et al. GASP. XV. A MUSE view of extreme ram-pressure stripping along the line of sight: physical properties of the jellyfish galaxy JO201. {\em MNRAS} {\bf 2019}, {\em 485}, 1157-1170

\bibitem[\protect\citeauthoryear{Benitez et al.}{2014}]{jpas}
Benitez, N. et al. 2014. J-PAS: The Javalambre-Physics of the Accelerated Universe Astrophysical Survey, arXiv:1403.5237


\bibitem[\protect\citeauthoryear{Bundy et al.}{2015}]{bundy15}
Bundy, K. et al 2015.
Overview of the SDSS-IV MaNGA Survey: Mapping nearby Galaxies at Apache Point Observatory,
{\em ApJ} {\bf 2015}, {\em 798}, 7

\bibitem[\protect\citeauthoryear{Crain et al.}{2015}]{eaglepaper_det}
Crain, R. et al. The EAGLE simulations of galaxy formation: calibration of subgrid physics and model variations. {\em MNRAS} {\bf 2015}, {\em 450}, 1937-1961.

\bibitem[\protect\citeauthoryear{Croom et al.}{2012}]{croom12}
Croom, R. et al. 2012. The Sydney-AAO Multi-object Integral field spectrograph {\em MNRAS} {\bf 2012}, {\em 421}, 872.


\bibitem[\protect\citeauthoryear{Dalla Vecchia \& Schaye}{2008}]{dallavecchiaschaye08}
Dalla Vecchia C., Schaye J., 2008, {\em MNRAS}, 387, 1431

\bibitem[\protect\citeauthoryear{Dalla Vecchia \& Schaye}{2012}]{dallavecchiaschaye12}
Dalla Vecchia C., Schaye J., 2012, {\em MNRAS}, 426, 140

\bibitem[\protect\citeauthoryear{Dolag et al.}{2009}]{dolag09}
Dolag K., Borgani S., Murante G., Springel V..
{\em MNRAS}, {\bf 2009}, {\em 399}, 497

\bibitem[\protect\citeauthoryear{Davies et al.}{2018}]{davies18}
Davies, L.J.M. et al. Deep Extragalactic VIsible Legacy Survey (DEVILS): motivation, design, and target catalogue. {\em MNRAS}, {\bf 2018}, {\em 480}, 768

\bibitem[\protect\citeauthoryear{Dawson et al.}{2013}]{boss13}
Dawson, K.~S. et al. The Baryon Oscillation Spectroscopic Survey of SDSS-III. {\em AJ} {\bf 2013}, 145, 10

\bibitem[\protect\citeauthoryear{Driver et al.}{2018}]{waves}
Driver, S. P., et al. The Wide Area VISTA Extra-galactic Survey (WAVES), arXiv1507.00676

\bibitem[\protect\citeauthoryear{EAGLE team}{2017}]{eaglePDR}
EAGLE team. The EAGLE simulations of galaxy formation: Public release of particle data, arXiv:1706.09899. 

\bibitem[\protect\citeauthoryear{Furlong et al.}{2015}]{furlong15} Furlong, M. et al. 2015. Evolution of galaxy stellar masses and star formation rates in the EAGLE simulations  {\em MNRAS} {\bf 2015}, {\em 450}, 4486.

\bibitem[\protect\citeauthoryear{Gunn \& Gott}{1972}]{gunngott72}
Gunn, J. E. \& Gott, J. R. 1972. On the infall of matter into clusters of galaxies and 
some effects on their evolution, {\em ApJ}, {\bf 1972}, 176, 1-19


\bibitem[\protect\citeauthoryear{Ivezic et al.}{2008}]{lsst}
Ivezic, Z. et al. 2008. Large Synoptic Survey Telescope: From Science Drivers To Reference Design. {\em Serbian Astronomical Journal} {\bf 2008}, {\em 176}, 1-13.

\bibitem[\protect\citeauthoryear{Jaff\'e et al.}{2015}]{jaffe15}
Jaff\'e, Y. et al. BUDHIES II: a phase-space view of H I gas stripping and star formation quenching in cluster galaxies. {\em MNRAS} {\bf 2015}, 448, 1715–1728


\bibitem[\protect\citeauthoryear{Jaff\'e et al.}{2018}]{yara18}
Jaff\'e, Y. et al. GASP. IX. Jellyfish galaxies in phase-space: an orbital study of intense ram-pressure stripping in clusters. {\em MNRAS} {\bf 2018}, 476, 4753-4764 (2018)

\bibitem[\protect\citeauthoryear{Kapferer et al.}{2009}]{kapferer09}
Kapferer, W., et al. 2009,
The effect of ram pressure on the star formation, mass distribution and morphology of galaxies, {\em A\&A} {\bf 2009}, 499, 87.

\bibitem[\protect\citeauthoryear{Kennicutt}{1998}]{kslaw} 
Kennicutt R. C., Jr, 1998. The Global Schmidt Law in Star-forming Galaxies
Kennicutt, {\em ApJ} {\bf 498}, {\em 541}, .


\bibitem[\protect\citeauthoryear{Koyama et al.}{2013}]{koyama13}
Koyama, Y., et al. 2013, On the evolution and environmental dependence of the star formation rate versus stellar mass relation since z $\sim$ 2. {\em MNRAS}, {\bf 2013}, {\em 434}, 423.

\bibitem[\protect\citeauthoryear{Lagos et al.}{2015}]{clau_gas}
Lagos C.~P., et al. 2015, Molecular hydrogen abundances of galaxies in the EAGLE simulations. {\em MNRAS}, {\bf 2015}, {\em 452}, 3815.

\bibitem[\protect\citeauthoryear{Lagos et al.}{2016}]
{clau_fp} Lagos C.~P., et al. 2016,
The Fundamental Plane of star formation in galaxies revealed by the EAGLE hydrodynamical simulations, 
{\em MNRAS}, {\bf 2016}, {\em 459}, 2632.

\bibitem[\protect\citeauthoryear{Lin et al.}{2014}]{lin14}
Lin, L. et al. 2014. The Pan-STARRS1 Medium-Deep Survey: The Role of Galaxy Group Environment in the Star Formation Rate versus Stellar Mass Relation and Quiescent Fraction out to z $\sim$ 0.8. {\em ApJ}, {\bf 2014}, {\em 782}, 33.

\bibitem[\protect\citeauthoryear{McAlpine et al.}{2016}]
{mcalpine16} McAlpine, S. et al. 2016. 
The EAGLE simulations of galaxy formation: Public release of halo and galaxy catalogues, {\em Astronomy and Computing}, {\bf 2016}, {\em 15}, 72-89.

\bibitem[\protect\citeauthoryear{McPartland et al.}{2016}]{mcpartland16}
McPartland, C.., 2016; Ebeling, H.; Roediger, E.; Blumenthal, K.  Jellyfish: The origin and distribution of extreme ram-pressure stripping events in massive galaxy clusters. {\em MNRAS} {\bf 2016}, {\em 455}, 2994--3008.

\bibitem[\protect\citeauthoryear{Moore et al. }{1996}]{moore96}
Moore, B. et al. Galaxy harassment and the evolution of clusters of galaxies. 
{\em Nature}, {\bf 1996}, 379, 613-616.
 
\bibitem[\protect\citeauthoryear{Muzzin et al.}{2012}]{muzzin}
Muzzin, A. et al. The Gemini Cluster Astrophysics Spectroscopic Survey (GCLASS): The Role of Environment and Self-regulation in Galaxy Evolution at z $\sim$ 1. 
{\em ApJ}, {\bf 2012}, {\em 746}, 188-213.

\bibitem[\protect\citeauthoryear{Pattarakijwanich et al.}{2016}]{pattarakijwanich16}
Pattarakijwanich, P.; Strauss, M. A.; Ho, S.; Ross, N. P. 
The Evolution of Post-Starburst Galaxies from $z\sim1$ to the Present. {\em AJ} {\bf 2016}, 833, 19. 

\bibitem[\protect\citeauthoryear{Peng et al.}{2010}]{peng10} Peng, Y-J. et al. 2010. Mass and environment as drivers of galaxy evolution in SDSS AND zCOSMOS and the origin of the Schechter function. {\em AJ } {\bf 2010}, {\em 721}, 193.

\bibitem[\protect\citeauthoryear{Poggianti et al.}{2016}]{poggianti16}
Poggianti, B. M. et al. Jellyfish galaxy candidates at low redshift. {\em AJ} {\bf 2016}, 151, 78-97.

\bibitem[\protect\citeauthoryear{Poggianti et al.}{2017}]{poggianti17}
Poggianti, B. M. et al. GASP. I. Gas Stripping Phenomena in Galaxies with MUSE. {\em ApJ} {\bf 2017}, 844, 48.

\bibitem[\protect\citeauthoryear{Rodr\'iguez et al. subm.}{2019}]{silvio}
Rodr\'iguez, S. et al. submitted to {\em MNRAS}.
Following the crumbs: Statistical effects of Ram Pressure in Galaxies. {\bf MN-19-2469-MJ}, {\em }.

\bibitem[\protect\citeauthoryear{Rosas-Guevara et al.}{2015}]{rosasguevara15}
Rosas-Guevara, Y. M. et al. 2015. The impact of angular momentum on black hole accretion rates in simulations of galaxy formation. {\em MNRAS} {\bf 454}, {\em 1038}.

\bibitem[\protect\citeauthoryear{Safarzadeh \&  Loeb}{2019}]{safarzadehloeb19}
Safarzadeh, M. and Loeb, L. Explaining the enhanced star formation rate of Jellyfish galaxies in galaxy
clusters. {\em MNRAS}, {\bf 2019}, {\em 486}, L26–L30.

\bibitem[\protect\citeauthoryear{Springel et al.}{2001}]{springel01}
Springel V., White S. D. M., Tormen G., Kauffmann G.
Populating a cluster of galaxies – I. Results at z = 0.
{\em MNRAS} {\bf 2001}, 328, 726.


\bibitem[\protect\citeauthoryear{Springel}{2005}]{springel05}
Springel, V. et al. The cosmological simulation code GADGET-2. {\em MNRAS} {\bf 2005}, 364, 1105.


\bibitem[\protect\citeauthoryear{Schaller et al.}{2015}]{schaller15}
Schaller, M. et al. 2015. The EAGLE simulations of galaxy formation: the importance of the hydrodynamics scheme.
{\em MNRAS} {\bf 2015}, {\em 454}, 2277.

\bibitem[\protect\citeauthoryear{Schaye}{2004}]{schaye04}
Schaye, J. 2004. Star Formation Thresholds and Galaxy Edges: Why and Where. {\em ApJ} {\bf 2004}, {\em 609}, 667.

\bibitem[\protect\citeauthoryear{Schaye et al.}{2015}]{eaglepaper}
Schaye, J. et al. The EAGLE project: simulating the evolution and assembly of galaxies and their environments. {\em MNRAS} {\bf 2015}, {\em 446}, 521-554.

\bibitem[\protect\citeauthoryear{Schaye \& Dalla Vecchia}{2008}]{schayedallavecchia08}
Schaye J., Dalla Vecchia C., 2008, {\em MNRAS}, 383, 1210


\bibitem[\protect\citeauthoryear{Shanks et al.}{2015}]{shanks15}
Shanks, T. et al. The VLT Survey Telescope ATLAS . {\em MNRAS} {\bf 2015}, {\em 451}, 4238


\bibitem[\protect\citeauthoryear{Smith et al.}{2010}]{smith10}
Smith, R.; Lucey, J.R.; Hammer, D.; Hornschemeier, A.E.; et al. Ultraviolet tails and trails in cluster galaxies: A sample of candidate gaseous stripping events in Coma.
{\em Mon. Notic. R. Astron. Soc.} {\bf 2015}, {\em 408}, 1417--1432.


\bibitem[\protect\citeauthoryear{Steinhauser et al.}{2012}]{steinhauser12}
Steinhauser, D. et al. Galaxies undergoing ram-pressure stripping: the influence
of the bulge on morphology and star formation rate,  {\em A\&A} {\bf 2012}, 544, A54.


\bibitem[\protect\citeauthoryear{Tissera et al.}{2017}]{patricia17}
Tissera, P. B. et al. Mild evolution of the stellar metallicity gradients of disc galaxies. {\em A\&A} {\bf 2017}, 604, 118.


\bibitem[\protect\citeauthoryear{Trayford et al.}{2015}]{trayford15}
Trayford, J. et al. 2015. Colours and luminosities of z = 0.1 galaxies in the EAGLE simulation {\em MNRAS} {\bf 2015 }, {\em 452}, 2879

\bibitem[\protect\citeauthoryear{Trayford et al.}{2017}]{trayford17}
Trayford, J. et al. 2017. Optical colours and spectral indices of z = 0.1 eagle galaxies with the 3D dust radiative transfer code skirt {\em MNRAS} {\bf 2017}, {\em 470}, 771

\bibitem[\protect\citeauthoryear{Trayford \& Schaye}{2019}]{trayford19}
Trayford, J. et al. 2019. Resolved galaxy scaling relations in the EAGLE simulation: star formation, metallicity, and stellar mass on kpc scales {\em MNRAS} {\bf 2019 }, {\em 485}, 5715
\bibitem[\protect\citeauthoryear{Troncoso Iribarren et al.}{2016}]{troncoso16b}
Troncoso-Iribarren, P. et al. 2016b.
Asymmetric Star Formation Efficiency Due to Ram Pressure Stripping.
{\em Galaxies} {\bf 2016}, {\em 4}, 77.

\bibitem[\protect\citeauthoryear{Yun et al.}{2019}]{yun19}
Yun, K., et al. 2019.
Jellyfish galaxies with the IllustrisTNG simulations - I. Gas-stripping phenomena in the full cosmological context.
{\em MNRAS} {\bf 2019}, {\em 483}, 1042.


\bibitem[\protect\citeauthoryear{Wendland}{1995}]{eaglekernel} 
Wendland, H. Piecewise polynomial, positive definite and compactly supported radial functions of minimal degree. {\em Advances in Computational Mathematics}, {\bf 1995}, {\em 4(1)}, 389-396.




\bibitem[\protect\citeauthoryear{Wiersma, Schaye \& Britton}{2009}]{wiersma09a}
Wiersma, Robert P. C.; Schaye, Joop; Smith, Britton D. 2009a. The effect of photoionization on the cooling rates of enriched, astrophysical plasmas.{\em MNRAS} {\bf 2009A}, {\em 393}, 99.

\bibitem[\protect\citeauthoryear{Wiersma et al.}{2009}]{wiersma09b}
Wiersma, R. P. C. et al. 2009b. Chemical enrichment in cosmological, smoothed particle hydrodynamics simulations. {\em MNRAS } {\bf 2009}, {\em 399}, 574.
\bibitem[\protect\citeauthoryear{Wright et al.}{2019}]{wright19}
Wright, R. et al. 2019. Quenching time-scales of galaxies in the EAGLE simulations. {\em MNRAS } {\bf 2019}, {\em 487}, 3740.
  

\end{thebibliography}




\appendix
\section{Dependency with cluster and galaxy properties}
\begin{figure*}
\includegraphics[width=2.\columnwidth, trim=4.5cm 14cm 3.5cm 3cm]{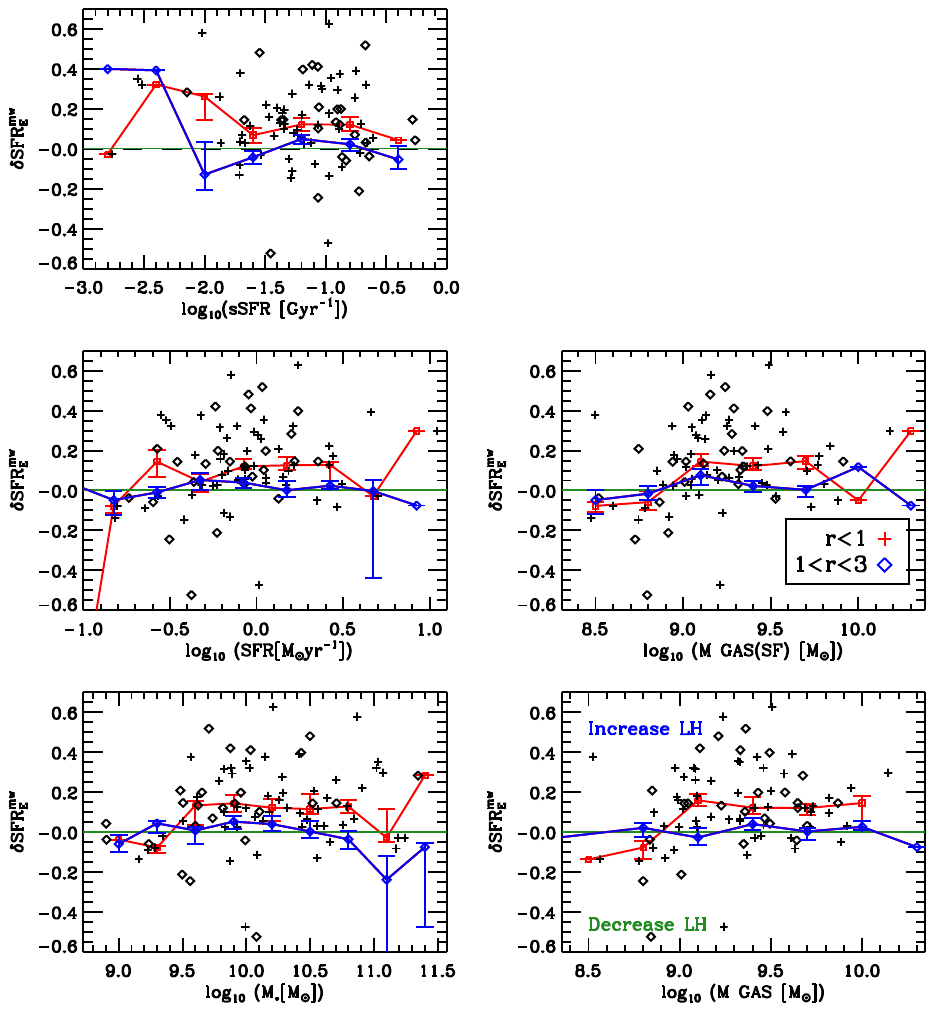}
\caption{Mass-weighted $\rm SFR$ enhancement, measured in the first case (see text), as function of galaxy intrinsic properties. From top to right, $\rm sSFR$, $\rm SFR$, mass of the star-forming gas, stellar mass, and total mass of gas.
In each panel, the solid green line shows the median value of the control sample.
Red, blue lines show the median of galaxies located within one and three virial radii, respectively.
The errors are the one sigma percentile.
Diamonds, crosses show satellites located within one virial radius, residing in relaxed and less relaxed cluster, respectively (see section \ref{sec_dynamical_state}).}
    \label{fig_sfrenh_gp}
\end{figure*}

\begin{figure*}
\includegraphics[width=2.\columnwidth, trim=4.5cm 14cm 3.5cm 3cm]{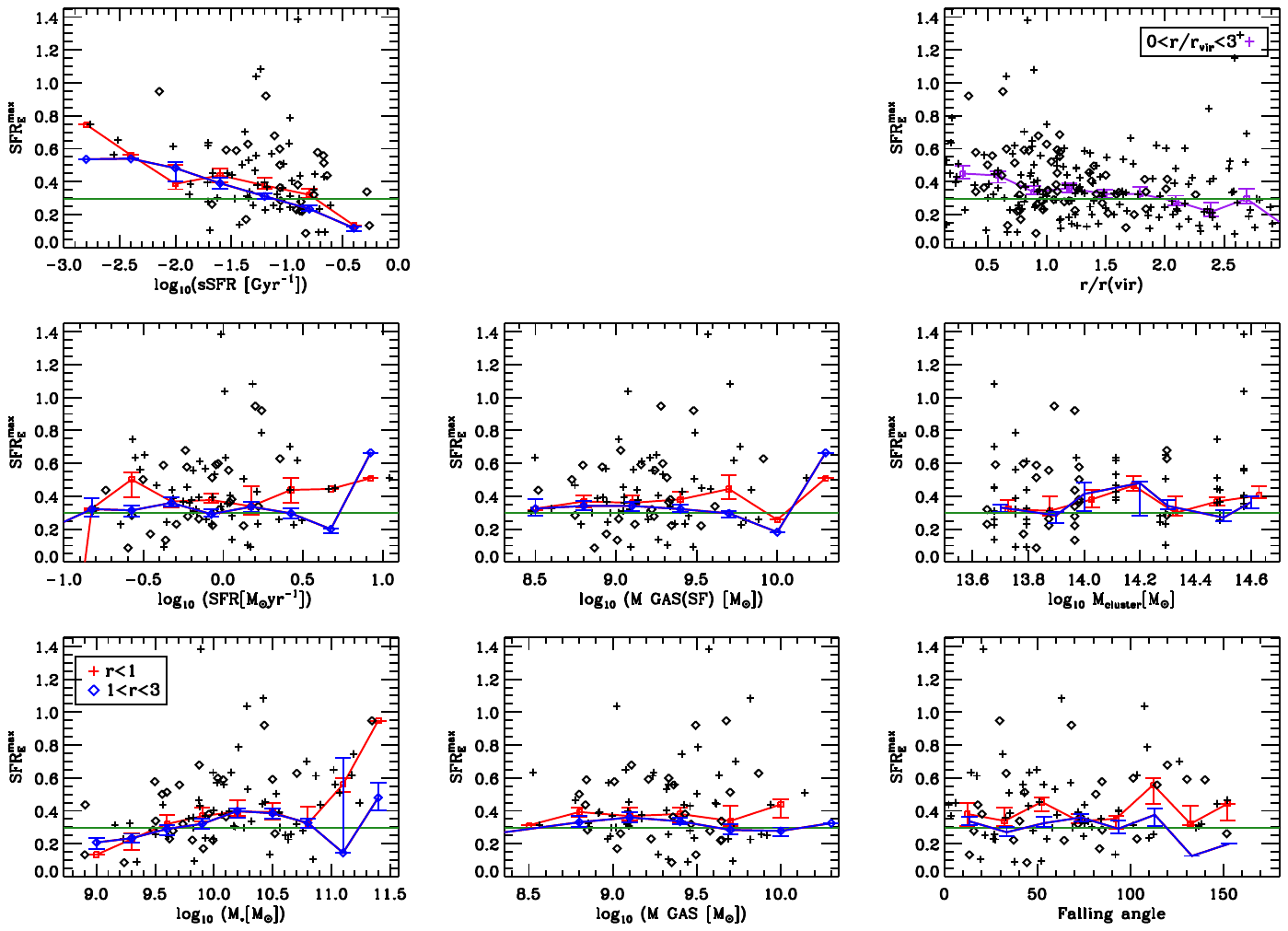}
\caption{
Maximum mass-weighted $\rm SFR$ enhancement (see text) as function of cluster and galaxy intrinsic properties.
{\it Left} and {\it middle} panels show the maximum $\rm SFR$ enhancement as a function of the galaxy intrinsic properties.
From top to right, $\rm sSFR$, $\rm SFR$, mass of the star-forming gas, stellar mass, and total mass of gas.
The {\it right} panels show the same quantity as a function of the cluster properties. From top to bottom, radial distance to the central galaxy normalized to the virial radius, cluster mass, and falling angle.
In each panel, the solid green line shows the median value of the control sample.
Diamonds show galaxies residing within one virial radius of relaxed cluster, while crosses in less relaxed clusters (see section \ref{sec_dynamical_state}).
Red and blue crosses indicate the median of the $\rm SFR_{E}^{max}$ at each bin for satellites within one and three virial radii, respectively.
The error bars show the one sigma percentile.}
    \label{fig_sfrmax_gp}
\end{figure*}

Here we analyze the dependency of the $\rm SFR$ and its maximum difference with cluster and galaxy properties.
We report the mass-weighted $\rm SFR$ because it can be directly related to observable quantities. 
In Figure \ref{fig_sfrenh_gp}, the mass-weighted $\rm SFR$ enhancement, measured in the first case (see text), as function of galaxy intrinsic properties is shown. From up to right, $\rm SFR$, star-forming gas, stellar mass, and total mass of gas. The red, blue lines show the median of galaxies located within one and three virial radii, respectively.
The Figure \ref{fig_sfrmax_gp} shows the maximum mass-weighted $\rm SFR$ difference, measured in the {\it observational case} (see text), as function of galaxy and cluster properties. In the left and middle panels, from up to right, $\rm SFR$, star-forming gas, stellar mass, and total mass of gas. In the right panel the $\rm SFR^{max}_{E}$ as a function of the distance to the central galaxy (top), cluster mass (middle), and falling angle (bottom). The red, blue lines show the median of galaxies located within one, and one and three virial radii, respectively.
In each panel, the solid green line shows the median value of the control sample.

\newpage
\begin{figure*}
\includegraphics[width=2.\columnwidth, trim=5cm 6.5cm 6cm 0.5cm]{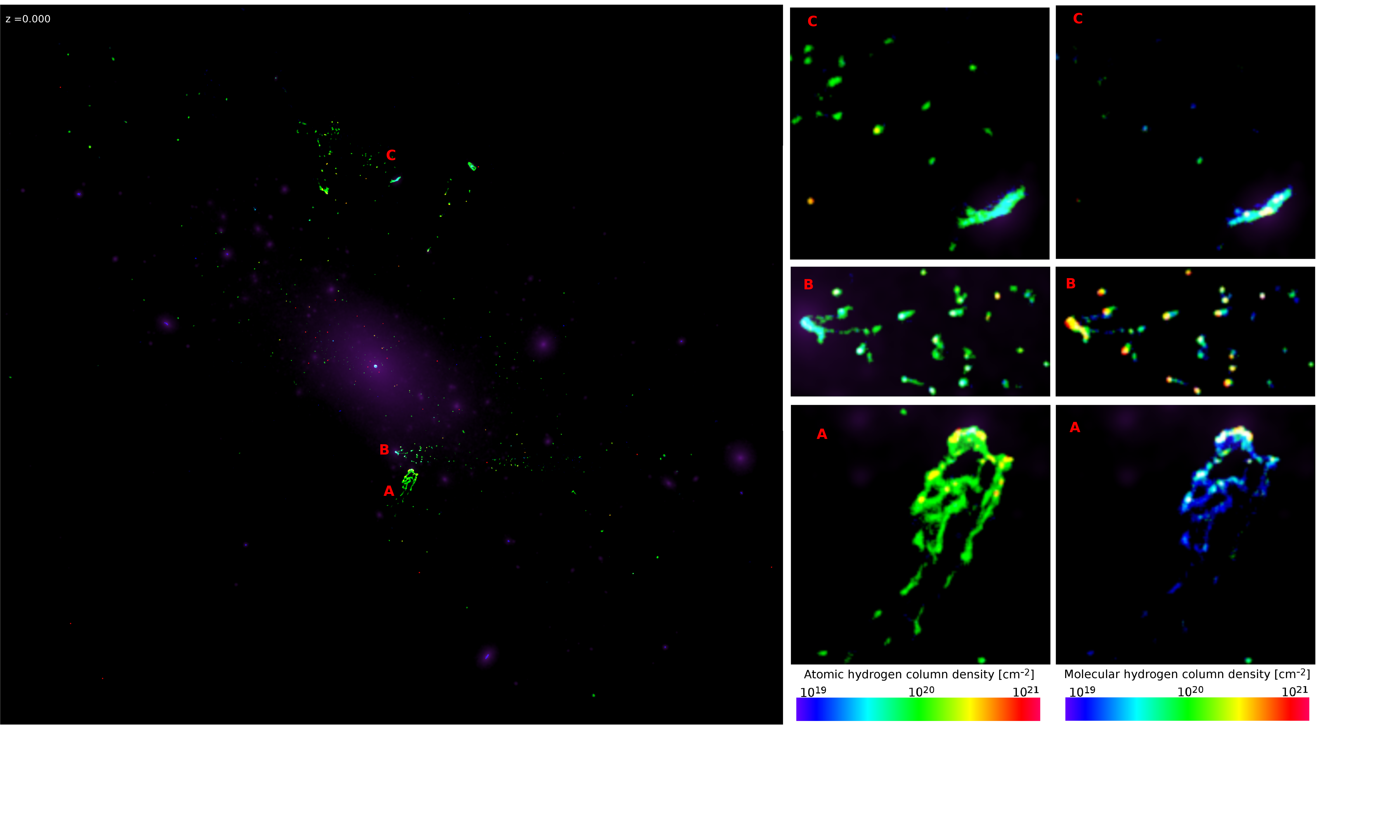}
\caption{Visualization of the halo of GroupNumber=8 in the EAGLE simulation. {\it Left:} Composite image of the cluster, the transparent purple represents the dark matter distribution, while the column density of the HI gas is shown in a color code scale. 
The HI and H2 maps are coloured by column density, according to the colour bars at the bottom, with column densities in units of $\rm cm^{-2}$. Particles are smoothed by 1 kpc in the NH2 and NHI maps. The separation between the gas components was performed according the recipes described in \protect\cite{clau_gas}. 
The field of view is 3$\times$3 $\rm Mpc$ and it is projected in the x-y plane of the simulation. {\it Right:} Example of Jellyfish galaxies in the same cluster. The bottom and top panels of HI and H2 maps have a size of 154 $\times$ 154 $\rm kpc^2$, while the middle one is 77 $\times$ 154 $\rm kpc^2$.}
\label{fig_jellyfishEAGLE}
\end{figure*}

\bsp	
\label{lastpage}
\end{document}